\documentclass[%
 reprint, 
 amsmath,amssymb,
 aps, physrev,
]{revtex4-2}

\usepackage{graphicx}
\usepackage{dcolumn}
\usepackage{bm}

\begin{document}

\preprint{APS/123-QED}

\title{\textbf{Post-Newtonian accelerations of a Mercury orbiter} 
}%

\author{Miriam Falletta}
\email{Contact author: miriam.falletta@phd.unipi.it}
\affiliation{%
 Department of Mathematics, University of Pisa, Largo Bruno Pontecorvo 5, 56127 Pisa, Italy
}%

\author{Gabriel Rodr\'{i}guez-Moris and Sergei A. Klioner}%

\affiliation{%
Lohrmann Observatory, Dresden Technical University, Mommsenstraße 13, 01062 Dresden, Germany
}%


\begin{abstract}
We investigate the relativistic modeling of spacecraft motion in Mercury's post-Newtonian local coordinates. This investigation is motivated by the fact that Mercury’s post-Newtonian gravitational field (as well as that of any other planet) admits an expansion in terms of multipole moments, which are most appropriately defined in the local reference system. The equations of motion in the Mercury-centric local frame include relativistic local perturbations, given by the Schwarzschild term, Lense-Thirring precession, and the acceleration due to the quadrupole moment, and relativistic third-body perturbations, which are the gravito-electric and gravito-magnetic accelerations, along with a coupling term between Mercury and other solar system bodies. The relativistic third-body perturbations are usually neglected in all practical applications. In this study, we analyze the magnitude of the post-Newtonian terms of the equations of motion formulated in the Mercury-centric frame, evaluating them along the trajectories of the two BepiColombo spacecrafts. Based on this analysis, we provide a practical approach for constructing a high-accuracy relativistic orbital model suitable for a Mercury orbiter.
\end{abstract}

\maketitle


\section{\label{sec: introduction} Introduction}
The level of accuracy required by current and upcoming space missions demands a degree of accuracy in dynamical modeling that goes beyond the Newtonian framework. In particular, a post-Newtonian level of accuracy is required both for the modeling of the motion of natural and artificial bodies in the solar system and for the propagation of electromagnetic waves. Relativistic modeling of dynamical phenomena within the solar system requires the use of several reference systems.
A consistent theory of astronomical reference systems at the first post-Newtonian approximation of general relativity was formulated by Brumberg and Kopeikin (see, e.g, \cite{kopeikin1988, Brumb_kop_proceedings, brumberg_kopeikin1989, brumbergkop1990, kopeikin1990, Klioner_voinov}) and by Damour, Soffel, and Xu \cite{DSX1, DSX2, DSX3, DSX4}. 

These two formalisms constitute the basis of the IAU 2000 Resolutions, which serve as practical guidelines in the fields of astrometry, celestial mechanics, and metrology \cite{IAU2000, Soffel2003}. The IAU 2000 Resolutions define the metric tensors and gravitational potentials of the Barycentric and Geocentric Celestial Reference Systems (BCRS and GCRS), as well as the coordinate transformations between them.

The BCRS, centered at the solar system barycenter and oriented as the International Celestial Reference System (ICRS) \cite{IAU2006B2}, is used to model the motion of both natural and artificial bodies within the solar system (e.g., planetary ephemerides, interplanetary spacecraft, etc). It also provides the appropriate framework for describing the positions and motions of distant galactic and extragalactic objects.
The GCRS, centered at the Earth’s barycenter, is used to describe phenomena occurring in the vicinity of the Earth (e.g., the motion of Earth satellites and Earth’s rotation). The GCRS is kinematically non-rotating with respect to the BCRS, that is, the linear part of the coordinate transformations between the spatial coordinates doesn’t contain a rotational component \cite{IAU2000, Soffel2003}. 

When describing physical processes near other planets, it might be necessary to introduce additional local systems, similarly to how the GCRS is defined. 
Note that in general relativity, no coordinate system is physically preferred, so physical phenomena within a given region of spacetime can be described using different arbitrary reference systems covering the same region. 
This means that, in principle, we can model the motion of an orbiter of a planet both in the local planetocentric frame and in the barycentric one. 

However, from a practical point of view, it is more convenient to adopt reference systems in which the mathematical formulation of physical laws is more straightforward. For instance, according to the Brumberg-Kopeikin and  Damour-Soffel-Xu formalisms, the multipole structure of the post-Newtonian gravitational field of a given body is adequately described only in the corresponding local reference system of that body.

The equations of motion for a spacecraft in the local reference frame can be derived by computing the geodesics of the local metric, as done in \cite{DSX4}. These equations include Newtonian terms due to the gravitational potential of the local body (monopole and multipole moments) and third-body perturbation, along with relativistic terms. The latter include relativistic local perturbations (for example, Schwarzschild term, Lense-Thirring precession, and the post-Newtonian acceleration due to the quadrupole moment of the local body), and relativistic third-body perturbations, which can be classified as terms independent of the velocity of the satellite (dubbed as “gravito-electric” below), those depending on the velocity of the satellite (dubbed as “gravito-magnetic”), and coupling terms between the local body and other solar system bodies. A general discussion of a number of these effects can be found in the recent review \cite{Iorio_2024}. In particular, since the local coordinates are kinematically non-rotating, the terms depending on the velocity of the satellite contain a relativistic Coriolis force due to the time-dependent rotation of the local coordinates with respect to the corresponding dynamically non-rotating reference system. Such a rotation has several components, often referred to as geodetic, Lense-Thirring, and Thomas precessions.

Except for the geodetic precession, relativistic third-body perturbations are usually ignored in all practical applications. This is partly because of the complexity of formulas expressing such accelerations, and partly because it is expected that they should have a very small magnitude. In contrast, spacecraft motion in the barycentric frame is typically modeled using the Einstein-Infeld-Hoffmann (EIH) equations \cite{EIH}, which include relativistic third-body terms, but assume that all bodies are point masses and neglect their extended structure and spin.
For the interested reader, a general discussion regarding post-Newtonian effects induced by the spin of a distant third body can be found in \cite{Iorio_2019, Iorio_2014}.

A first numerical investigation of relativistic external accelerations was carried out in \cite{Roh}, focusing on near-Earth satellites. In that study, the accelerations were computed using a linear expansion in terms of the satellite's geocentric distance, as derived in \cite{brumberg_kopeikin1989}. Note that such an expansion is suitable for near satellites, while for high-flying satellites, more complete expressions, such as the closed-form formulas reported in \cite{DSX4}, are preferable. 

In this study, starting from the basic principles of the IAU 2000 relativistic framework, we investigate the full post-Newtonian equations of motion for Mercury orbiters in Mercury's local coordinates, motivated by the fact that the acceleration due to the multipole structure of Mercury’s post-Newtonian gravitational field can be most conveniently described in the local reference system of the planet. Mercury, being the innermost planet of the solar system, offers a gravitational environment where relativistic perturbations from external bodies (mainly the Sun) can be expected to be more significant.
Starting from the foundations of the theory of astronomical reference systems, we present the construction of Mercury's local coordinates at the first post-Newtonian approximation of general relativity and discuss the order of magnitude of the various terms appearing in the equations of motion within this frame.  The analysis is conducted by numerically evaluating the relativistic accelerations along the orbits of the two BepiColombo spacecrafts.

BepiColombo is a mission to explore Mercury, jointly developed by the European Space Agency (ESA) and the Japan Aerospace Exploration Agency (JAXA) \cite{benkhoff}. Launched in October 2018, it is scheduled for Mercury orbit insertion in November 2026, with science operations beginning in April 2027. The nominal mission duration is one year, with a potential one-year extension. It consists of two spacecraft: the Mercury Planetary Orbiter (MPO) and the Mercury Magnetospheric Orbiter (MMO, since 2018 called Mio), which will be placed in distinct orbits around Mercury. The mission aims to study Mercury’s surface, internal structure, and magnetosphere. A key component of the scientific payload is the Mercury Orbiter Radioscience Experiment (MORE), which will provide highly accurate radio tracking data of the MPO to determine Mercury’s gravity field and rotational dynamics \cite{iess}, and to test general relativity by estimating the parameterized post-Newtonian (PPN) parameters \cite{Milani2002}.

The PPN formalism is the standard framework for testing metric theories of gravity, including general relativity, at the first post-Newtonian approximation \cite{willPPN}.  Consequently, developing a theory of astronomical reference systems in the PPN framework has become necessary. In \cite{relcelmechPPN}, the construction of both global and local PPN reference systems with PPN parameters $\gamma$ and $\beta$ is presented,  along with the derivation of the spacetime transformations between these systems and of the equations of motion for a spacecraft in the local PPN reference system. While one of the main objectives of the BepiColombo mission is to estimate the PPN parameters, the present work assumes general relativity as the underlying theory, given the tight experimental bounds that constrain the PPN parameters very close to their general relativistic values \cite{will2014}.

This paper is organized as follows. In Section~\ref{sec: spacetime reference frames}, we recall some basic results on the theory of astronomical reference frames in the first post-Newtonian approximation of general relativity, focusing in particular on the space-time coordinate transformations and the scaling of astronomical quantities. The construction of a Mercury-centric local reference system is also presented. 
Section~\ref{sec: numerical experiments} presents the results of the numerical evaluation of the various accelerations composing the equations of motion in the Mercury-centric local reference system along the MPO's and Mio's orbits.  Some concluding remarks are drawn in Section~\ref{sec: conclusions}. 
Finally, the equations of motion for a spacecraft in the Mercury-centric local reference system are presented in Appendix~\ref{sec: eqs of motion}, following the derivation given in \cite{relcelmechPPN} applied to the specific case of general relativity.

\section{Space-time reference frames}
\label{sec: spacetime reference frames}
\subsection{Notation}
The adopted notation follows that of \cite{DSX4} and \cite{relcelmechPPN}. We label the bodies of the $ N$-body system (i.e., the solar system) with capital letters. We use the letter M to refer to Mercury, and the letter X to refer to a generic body of the system. 
We denote as $G$ the gravitational constant, $c$ indicates the speed of light, and $M_\text{X}$ designates the mass of a generic body $\text{X}$.

Greek indices $\alpha, \beta, \dots, \mu, \nu, \dots$ run from 0 to 3 and refer to space-time components. Latin indices $a, b, \dots, i, j, \dots$ run from 1 to 3 and refer to space components only.
 
Coordinates in the global reference system are indicated by $x^\mu=(ct, x^i)$, where $t=\text{TCB}$ and Greek and Latin indices are taken from the second part of the two alphabets.
Coordinates in a generic local reference system are denoted using capital letters, as $X^\alpha= (cT, X^a)$ with Greek and Latin indices taken from the first part of the two alphabets. We denote as $T=\text{TCX}$ the time coordinate of the local reference system referred to body X. 

Repeated indices indicate the Einstein summation convention, e.g., $x^i y^i = x^1 y^1 + x^2 y^2 + x^3 y^3$. We denote with $\delta^{ij}=\text{diag}(1,1,1)$ the Kroenecker delta and with $ \varepsilon_{ijk}$ the fully antisymmetric Levi-Civita symbol ($ \varepsilon_{123}=+1$). Parentheses surrounding a group of indices of any tensor denote symmetrization over these indices (e.g. $T_{(ij)}= (T_{ij} + T_{ji})/2$), while brackets enclosing a group of indices indicate anti-symmetrization (e.g. $T_{[ij]}= (T_{ij} - T_{ji})/2$). Three-dimensional space vectors are denoted as $\bm{v}=v^i$. A dot denotes the Euclidean scalar product $\bm{v}\cdot \bm{w}= v^i w^i$, and the usual Euclidean norm of a vector is indicated as $|\bm{v}|=\sqrt{\bm{v}\cdot \bm{v}}$.

A comma before an index designates a partial derivative, i.e. $A_{,\mu}= \partial A(t, \bm{x}) / \partial x^{\mu}$, $A_{, i}= \partial A (t, \bm{x})/ \partial x^{i}$, $A_{,\alpha}= \partial A(T, \bm{X}) / \partial X^{\alpha}$, $A_{, a}= \partial A(T, \bm{X}) / \partial X^{a}$. Partial derivatives with respect to time are denoted as $A_{,t}= \partial A(t, \bm{x}) / \partial t$, $A_{,T}= \partial A (T, \bm{X})/ \partial T$. In what follows, we will often deal with functions evaluated in the world line of the local body, which will be shortly denoted as $A(\bm{x}_\text{X})$ (i.e. $A(\bm{x}_\text{X})= A(t, \bm{x}_\text{X}(t))$). Moreover, for such functions, the notation $A_{,i}(\bm{x}_\text{X})$ denotes $ \partial A(t, \bm{x})/\partial x^i$ evaluated at $\bm{x}=\bm{x}_\text{X}$. Analogously, the notation $A_{,t}(\bm{x}_\text{X})$ denotes $ \partial A(t, \bm{x})/\partial t$ evaluated at $\bm{x}=\bm{x}_\text{X}$, with similar relations holding for partial derivatives with respect to local coordinates.

\subsection{Space-time transformations}

The space-time transformations from the global coordinates $(t, x^i)$ to the coordinates $(T, X^a)$ of the local reference system around a generic body X read \cite{IAU2000}
\begin{eqnarray}
    \nonumber
    T & =&t- \frac{1}{c^2}\left(A(t) + v_\text{X}^ir_\text{X}^i \right)+ \frac{1}{c^4} \left(B(t) + B^i(t)r_\text{X}^i \right.\\
    \label{eq: time transformation}
    & &\left. + B^{ij}(t)r_\text{X}^ir_\text{X}^j +C(t, \bm{x}) \right) + O(c^{-5}),\\
    \nonumber
    {X^a} & =& \delta_i^a\left[
    {r}_\text{X}^i+\dfrac{1}{c^2}\left(
\dfrac{1}{2}\,{v}_\text{X}^i{v}^k_\text{X}{r}^k_\text{X}+\overline{w}(\bm{x}_\text{X}) {r}_\text{X}^i+ {r}_\text{X}^i{a}_\text{X}^k{r}^k_\text{X} \right. \right.\\
\label{eq: space transformation}
 & &\left. \left. -\dfrac{1}{2}{a}_\text{X}^i|\bm{r}_\text{X}|^2
\right)\right]+O(c^{-4})\,.
\end{eqnarray}
 where $\bm{r}_\text{X}(t, \bm{x})=\bm{x}-\bm{x}_\text{X}(t)$, $\bm{x}_\text{X}(t)$ are the coordinates of the origin of the local reference system in the global one, and $\bm{v}_\text{X}(t)=\dfrac{d}{dt}\,\bm{x}_\text{X}(t)$ and
$\bm{a}_\text{X}(t)=\dfrac{d}{dt}\,\bm{v}_\text{X}(t)$ are its velocity and acceleration, respectively. Here
\begin{eqnarray*}
    \frac{d}{dt}{A}(t)&=& \frac{1}{2}|\bm{v}_\text{X}|^2 + \overline{w}(\bm{x}_\text{X}),\\
    \frac{d}{dt}{B}(t)&=&-\frac{1}{8}|\bm{v}_\text{X}|^4-\frac{3}{2}|\bm{v}_\text{X}|^2\overline{w}(\bm{x}_\text{X})+ 4v_\text{X}^i\overline{w}^i(\bm{x}_\text{X})\\
    && +\frac{1}{2}\overline{w}^2(\bm{x}_\text{X}), \\
    B^i(t) & = & -\frac{1}{2}|\bm{v}_\text{X}|^2 v_\text{X}^i +4\overline{w}^i(\bm{x}_\text{X})-3 v_\text{X}^i\overline{w}(\bm{x}_\text{X}), \\
    B^{ij}(t)&= &- v_\text{X}^i\delta_{aj}Q^a+ 2\frac{\partial}{\partial x^j}\overline{w}^i(\bm{x}_\text{X})-v_\text{X}^i\frac{\partial}{\partial x^j}\overline{w}(\bm{x}_\text{X}) \\
    && + \frac{1}{2}\delta^{ij}\frac{d}{dt}{\overline{w}}(\bm{x}_\text{X}), \\
    C(t, \bm{x})&=&-\frac{1}{10}|\bm{r}_\text{X}|^2\left(r^i_\text{X}\frac{d}{dt}{a}^i_\text{X}\right).
\end{eqnarray*}
Moreover, at the Newtonian level, 
\begin{equation*}
    Q^a= \delta_{ai}\left[\frac{\partial}{\partial x^i} \overline{w}(\bm{x}_\text{X}) -a^i_\text{X}\right], 
\end{equation*}
and $\overline{w}$ and $\overline{w}^i$ are the global scalar and vector external potentials, given by 
\begin{equation*}
    \overline{w}(t, \bm{x}) = \sum_{\text{A}\neq \text{X}} w_\text{A}(t, \bm{x}), \quad \overline{w}^i(t, \bm{x}) = \sum_{\text{A}\neq \text{X}} w_\text{A}^i(t, \bm{x}).
\end{equation*}
If the external bodies are approximated as mass monopoles, the potentials $w_\text{A}$, $w^i_\text{A}$ read 
\begin{align*}
 w_\text{A}(t, \bm{x})& =\frac{GM_\text{A}}{|\bm{r}_\text{A}|} -\frac{1}{c^2}\Delta_\text{A}(t, \bm{x}), \\ 
    {w}_\text{A}^i(t, \bm{x}) & =  \frac{GM_\text{A}}{|\bm{r}_\text{A}|}v_\text{A}^i,  
\end{align*}
where, denoting as $\bm{r}_\text{BA}(t)=\bm{x}_\text{B}(t)-\bm{x}_\text{A}(t)$, the function $\Delta_\text{A}(t, \bm{x})$ is given by: 
\begin{eqnarray*}
    \Delta_\text{A}(t, \bm{x})&=&\frac{GM_\text{A}}{|\bm{r}_\text{A}|}\left[-2|\bm{v}_\text{A}|^2+\sum_{\text{B}\neq\text{A}}\frac{GM_\text{B}}{|\bm{r}_\text{BA}|} \right. \\ 
    & & \left. + \frac{1}{2}\left( \frac{(r_\text{A}^k v_\text{A}^k)^2}{|\bm{r}_\text{A}|^2} + r_\text{A}^k a_\text{A}^k\right)\right].
\end{eqnarray*}
 
The time transformation shown in Eq.~(\ref{eq: time transformation}) at the center of the local body (that is, at $\bm{x}=\bm{x}_\text{X}(t)$) can be rewritten as an ordinary differential equation. 
Denoting as $T_\text{center}$ the local coordinate time computed at the center of the local body, the differential equation expressing the time transformation at the center of the local body reads: 
\begin{equation}
    \label{eq: time transformation geocenter}
    \frac{dT_{\text{center}}}{dt} = 1+F(t), 
\end{equation}
where $F(t)=-\dfrac{1}{c^2}\dfrac{d}{dt}{A}(t) +\dfrac{1}{c^4}\dfrac{d}{dt}{B}(t) + O(c^{-5}) $. 
Equivalently, we can define a function $\Delta t(t)$ such that 
\begin{equation}
\label{eq: def of deltat(t)}
    T_\text{center}=t+ \Delta t(t),
\end{equation}
 satisfying the following differential equation:
\begin{equation}
\label{eq: ode delta t}
    \frac{d\Delta t}{dt} =F(t).
\end{equation}

\subsection{Scalings}
The time coordinate in the BCRS is called Barycentric Coordinate Time (TCB), while in the GCRS it is called Geocentric Coordinate Time (TCG). 
However, for reasons of practical utility, two additional time scales, Barycentric Dynamical Time (TDB) and Terrestrial Time (TT), are used. TDB and TT are defined as linear functions of TCB and TCG, respectively. 
In particular, it holds that
\begin{equation}
\label{eq: TDB}
\text{TDB} = \text{TCB} - L_B \times (\text{JD}_\text{TCB} -T_0)\times 86400 + \text{TDB}_0,
\end{equation}
where $T_0=2443144.5003725$, and $L_\text{B}=1.550519768\times10^{-8}$ and $\text{TDB}_0=-6.55\times10^{-5}$ s are defining constants whose values are chosen by the IAU 2006 Resolution B3 \cite{IAU2006B3}. $\text{JD}_\text{TCB}$ is the TCB Julian date, whose value was set to be equal to $T_0$ for the event
1 January 1977 00h 00m 32.184s TT at the geocenter by the IAU. The definition of TDB (Eq.~(\ref{eq: TDB})) ensures that TDB, evaluated at the geocenter, remains as close as possible to TT. As described in \cite{klioner_scaling}, the choice of the scaled versions of the time coordinates implies the adoption of the scaled spatial coordinates and scaled mass parameters, where the mass parameter for a body is defined as the product of its mass and the gravitational constant $G$. 
In particular, denoting as $\bm{x}^*$ the scaled spatial coordinates and as $\mu^*$ the scaled mass parameters for each body, it holds that
\begin{eqnarray}
    \label{eq: scaling spatial coordinate TDB}
    \bm{x}^*&=&(1-L_\text{B})\bm{x}, \\
    \label{eq: scaling mass parameter TDB}
    \mu^*&=&(1-L_\text{B})\mu.
\end{eqnarray}
The quantities $\mu^*$, $\bm{x}^*$ and $t^*=\text{TDB}$ are called TDB-compatible, while $\mu$, $\bm{x}$ and $t=\text{TCB}$ are TCB-compatible quantities \cite{klioner_scaling}. Note that $\mu$ is the physical mass parameter, while $\mu^*$ has no physical meaning, being just an artificial parameter due to the scaling of coordinates. 

Let us now introduce the scaled time coordinate $T^{**}={\text{TX}}$, such that 
\begin{equation}
\label{eq: TDX}
\text{TX} = \text{TCX} - L_\text{X}\times(\text{JD}_\text{TCX} - T_0)\times 86400
\end{equation}
where $\text{JD}_\text{TCX}$ is the TCX Julian date, and $L_\text{X}$ is a defining constant. For instance, in the case of the Earth, one has $L_\text{G}=6.969290134\times10^{-10}$ \cite{IAU2000}. Note that the value of the constant $L_\text{G}$ has been chosen so that, at current clock accuracies, the mean rate of TT matches the mean rate of the proper time of an observer situated on the rotating geoid \cite{klioner_scaling}. In contrast, the definition of $L_\text{X}$ for bodies other than the Earth is not necessarily based on this consideration.

Analogously, one can introduce the scaled spatial coordinates $\bm{X}^{**}$ and $\mu^{**}$, as 
\begin{eqnarray}
\label{eq: scaling spatial coordinates}
   \bm{X}^{**}&=&(1-L_\text{X})\bm{X}, \\
   \label{eq: scaling mass parameter}
    \mu^{**}&=&(1-L_\text{X})\mu,
\end{eqnarray}
where, as usual, $\mu^{**}$, $\bm{X}^{**}$ and $T^{**}$ are {TX}-compatible quantities, while $\mu$, $\bm{X}$ and $T$ are TCX-compatible. As stated above, the physical mass parameter of a given body corresponds to $\mu$, and it has the same value in the BCRS, in the GCRS, and in each GCRS-like planetocentric reference system. Only its scaled TDB-compatible, TT-compatible, or TX-compatible values differ. 

Similarly to Eq.~(\ref{eq: def of deltat(t)}), the relation between $t^*=\text{TDB}$ and $T^{**}={\text{TX}}$ at the center of the local body reads 
\begin{equation}
    \label{eq: TDX-TDB}
    T^{**}_\text{center}=t^* +\Delta t^*(t^*). 
\end{equation}
The function $\Delta t^*(t^*)$ satisfies the following differential equation:
\begin{equation}
\label{eq: ODE TDB TDX}
    \frac{d\Delta t^*}{dt^*}= C_1 + C_2 \frac{d\Delta t}{dt},
\end{equation}
where the derivative ${d\Delta t}/{dt}$ is defined in Eq.~(\ref{eq: ode delta t}) and must be expressed as a function of $t^*$. Moreover, the constants $C_1$ and $C_2$ in Eq.~(\ref{eq: ODE TDB TDX}) are given by
\begin{eqnarray*}
    C_1 &=& \frac{L_\text{B}-L_\text{X}}{1-L_\text{B}}, \\
    C_2 &=&\frac{1- L_\text{X}}{1-L_\text{B}}=1+C_1.
\end{eqnarray*}
Therefore, given a TDB time, in order to compute the {TX} time, one can numerically solve the differential equation~(\ref{eq: ODE TDB TDX}), starting from an adequately chosen initial condition \cite{klioner_gaiatime}. The initial condition follows from Eqs.~(\ref{eq: TDB}) and (\ref{eq: TDX}), i.e., at the Julian Date of the 1st of January 1977, 00:00:32.184 s TX ($\text{JD}_{\text{TX}}=2443144.5003725$), one has $\text{JD}_{\text{TDB}} = 2443144.5003725- 6.55\times 10^{-5}/86400$. 

Subsequently, to get the time {TX} at the desired location, one should add the location-dependent terms in Eq.~(\ref{eq: time transformation}), computed using TDB-compatible quantities. 

\subsection{Transformations of accelerations}
\label{sec: transformations of accelerations}
Given the position $\bm{x}_\text{s}(t)$ of a spacecraft at time $t$ in the barycentric reference system, the position vector $\bm{X}_\text{s}(T)$ in the local planetocentric system can be computed from Eq.~(\ref{eq: space transformation}) and reads
\begin{eqnarray}
\nonumber
\bm{X}_\text{s} &=& \bm{r}_\text{s{X}}+\dfrac{1}{c^2}\left(
\dfrac{1}{2}\bm{v}_\text{X}(\bm{v}_\text{X}\cdot\bm{r}_\text{s{X}})+\overline{w}(\bm{x}_\text{X}) \bm{r}_\text{s{X}} \right. \\
\label{eq: transf pos spacecraft}
& & \left. +\bm{r}_\text{s{X}}(\bm{a}_\text{X}\cdot\bm{r}_\text{s{X}})
 -\dfrac{1}{2}\bm{a}_\text{X}\,|\bm{r}_\text{s{X}}|^2
\right)+O(c^{-4}),
\end{eqnarray}
where $\bm{r}_\text{sX}(t)=\bm{x}_\text{s}(t)- \bm{x}_\text{X}(t)$, and $T$ is given by the time transformation evaluated at $(t,\bm{x})=(t, \bm{x}_\text{s}(t))$. 

The barycentric velocity can be transformed into the corresponding velocity in the local planetocentric frame by taking the time derivative of Eq.~(\ref{eq: transf pos spacecraft}) and using the time transformation. The velocity vector $\dot{\bm{X}}_\text{s}(T)= d \bm{X}_\text{s}(T) / dT$ in the local planetocentric system reads
\begin{eqnarray}
    \nonumber
    \frac{d}{dT} \bm{X}_\text{s}& = &\delta\bm{v}_\text{s} + \frac{1}{c^2} \left[ \delta\bm{v}_\text{s} \left(\frac{1}{2}|\bm{v}_\text{X}|^2 + 2(\overline{w}(\bm{x}_\text{X}) +\bm{a}_\text{X} \cdot \bm{r}_{\text{s}\text{X}}) \right. \right. \\ 
    \nonumber 
    && \left. \left. + \bm{v}_\text{X} \cdot \delta\bm{v}_\text{s} \right)+ \frac{1}{2} \bm{v}_\text{X} (\bm{v}_\text{X} \cdot \delta\bm{v}_\text{s} ) + \bm{r}_{\text{s}\text{X}} (\bm{a}_\text{X} \cdot \delta\bm{v}_\text{s}) \right. \\
   \nonumber
   && \left. - \bm{a}_\text{X} (\bm{r}_{\text{s}\text{X}} \cdot \delta\bm{v}_\text{s})  + \frac{1}{2} \bm{a}_\text{X} (\bm{v}_\text{X} \cdot \bm{r}_{\text{s}\text{X}}) \right. \\ 
    \nonumber
    && \left.  + \frac{1}{2} \bm{v}_\text{X} (\bm{a}_\text{X}\cdot \bm{r}_{\text{s}\text{X}}) + \dot{\overline{w}}(\bm{x}_\text{X}) \bm{r}_{\text{s}\text{X}}  +  \bm{r}_{\text{s}\text{X}} (\dot{\bm{a}}_\text{X}\cdot\bm{r}_{\text{s}\text{X}} ) \right. \\
    \label{eq: transf velocity} 
    && \left. - \frac{1}{2} \dot{\bm{a}}_\text{X} |\bm{r}_{\text{s}\text{X}}|^2 \right] + O(c^{-4}),
\end{eqnarray}
where $\delta\bm{v}_\text{s}(t) = \dfrac{d}{dt}{\bm{x}}_\text{s} (t)- \bm{v}_\text{X}(t)$, $\dot{\overline{w}}(\bm{x}_\text{X})= \dfrac{d}{dt}\overline{w}(\bm{x}_\text{X})$ and $\dot{\bm{a}}_\text{X}(t)= \dfrac{d}{dt}\bm{a}_\text{X}(t)$.

Finally, retaking the time derivative and again using the time transformation, the acceleration vector $\ddot{\bm{X}}_\text{s}(T)= d^2 \bm{X}_\text{s}(T) / dT^2$ in the local planetocentric system can be computed as
\begin{eqnarray}
    \nonumber
  \hspace*{-0.7em} 
  \frac{d^2}{dT^2} \bm{X}_\text{s} & & =   \delta\bm{a}_\text{s} \\
     \nonumber
    && + \frac{1}{c^2} \left[ \delta\bm{a}_\text{s} \left( |\bm{v}_\text{X}|^2 + 3\overline{w}(\bm{x}_\text{X})+ 3\bm{a}_\text{X} \cdot \bm{r}_{\text{s}\text{X}} + 2 \delta\bm{v}_\text{s} \cdot \bm{v}_\text{X} \right)  \right. \\ 
     \nonumber
    &&    + \delta\bm{v}_\text{s} \left( 4 \bm{a}_\text{X}\cdot\delta\bm{v}_\text{s} + 3 \dot{\bm{a}}_\text{X} \cdot \bm{r}_{\text{s}\text{X}} \right. \\
    \nonumber
    && \left. + 3 \dot{\overline{w}}(\bm{x}_\text{X})  +  \bm{a}_\text{X}\cdot\bm{v}_\text{X} +\delta\bm{a}_\text{s} \cdot\bm{v}_\text{X} \right) \\
    \nonumber
     && + \bm{v}_\text{X}\left(\bm{a}_\text{X} \cdot  \delta\bm{v}_\text{s} + \frac{1}{2}\delta \bm{a}_\text{s} \cdot \bm{v}_\text{X}  +  \frac{1}{2} \dot{\bm{a}}_\text{X} \cdot  \bm{r}_{\text{s}\text{X}} \right) \\
     \nonumber
     && + \bm{a}_\text{X} \left( \bm{v}_\text{X} \cdot \delta\bm{v}_\text{s} + \bm{a}_\text{X} \cdot \bm{r}_{\text{s}\text{X}} -  |\delta\bm{v}_\text{s}|^2 - \delta\bm{a}_\text{s} \cdot \bm{r}_{\text{s}\text{X}}\right) \nonumber\\ 
     && + \bm{r}_{\text{s}\text{X}} \left(  \ddot{\overline{w}}(\bm{x}_\text{X}) +  \ddot{\bm{a}}_\text{X}\cdot \bm{r}_{\text{s}\text{X}}+ 2 \delta\bm{v}_\text{s} \cdot\dot{\bm{a}}_\text{X} + \delta\bm{a}_\text{s} \cdot{\bm{a}}_\text{X}  \right) \nonumber  \\
     \nonumber
    & & + \dot{\bm{a}}_\text{X} \left(\frac{1}{2}\bm{v}_\text{X} \cdot \bm{r}_{\text{s}\text{X}} - 2 \delta\bm{v}_\text{s} \cdot \bm{r}_{\text{s}\text{X}} \right)  \\
    \label{eq: transf acc}
    && \left. - \frac{1}{2}\ddot{\bm{a}}_\text{X} |\bm{r}_{\text{s}\text{X}}|^2\right] + O(c^{-4}),
\end{eqnarray}
where $\delta\bm{a}_\text{s}(t) = \dfrac{d^2}{dt^2}{\bm{x}}_\text{s}(t) - \bm{a}_\text{X}(t) $,  $\ddot{\overline{w}}(\bm{x}_\text{X})= \dfrac{d^2}{dt^2}\overline{w}(\bm{x}_\text{X})$, and $\ddot{\bm{a}}_\text{X}(t)= \dfrac{d^2}{dt^2}\bm{a}_\text{X}(t)$.

Since Eqs.~(\ref{eq: transf pos spacecraft})--(\ref{eq: transf acc}) are derived from the transformation of spatial coordinates (Eq.~(\ref{eq: space transformation})), the quantities involved in the above formulas should be TCB-compatible on the right-hand side, and TCX-compatible on the left-hand side. 

Eq.~(\ref{eq: transf acc}) expresses the relation between the acceleration of the spacecraft $\ddot{\bm{x}}_\text{s}(t)=d^2\bm{x}_\text{s}(t)/dt^2$ in the barycentric frame and the acceleration $\ddot{\bm{X}}_\text{s}(T)=d^2\bm{X}_\text{s}(T)/dT^2 $ in the local planetocentric frame. In the case where all planets are mass monopoles, the barycentric acceleration of the orbiter $\ddot{\bm{x}}_\text{s}$ in the first post-Newtonian approximation is given by the EIH equations \cite{EIH} with the mass of the satellite being neglected:  
\begin{eqnarray}
\nonumber
\frac{d^2}{dt^2}{\bm{x}}_\text{s} &= & 
- \sum\limits_{\text{A}}{GM_\text{A}\frac{\bm{r}_{\text{s}\text{A}}}{|\bm{r}_{\text{s}\text{A}}|^3}} 
 +\frac{1}{c^2}\sum\limits_{\text{A}}GM_\text{A}\frac{\bm{r}_{\text{s}\text{A}}}{|\bm{r}_{\text{s}\text{A}}|^3} \\
 \nonumber
&& \times \left\{  
\sum\limits_{ \text{B}\ne \text{A}}{ \frac{GM_\text{B}}{|\bm{r}_{\text{A}\text{B}}|} }  \right.
-2\dot{\bm{x}}_\text{A}\cdot \dot{\bm{x}}_\text{A} 
- \dot{\bm{x}}_\text{s}\cdot\dot{\bm{x}}_\text{s} \\
\nonumber
&& +4\dot{\bm{x}}_\text{s}\cdot\dot{\bm{x}}_\text{A} + 4\sum\limits_{\text{B}}{\frac{GM_\text{B}}{|\bm{r}_{\text{s}\text{B}}|}}  
+ \frac{3}{2}\frac{{\left(\bm{r}_{\text{s}\text{A}}\cdot\dot{\bm{x}}_\text{A}\right)}^2}{|\bm{r}_{s\text{A}}|^2} \\ 
\label{EIH-eqm}
&& \left. \qquad \qquad - \frac{1}{2}\sum\limits_{ \text{B}\ne \text{A}} {GM_\text{B}\frac{\bm{r}_{\text{s}\text{A}}\cdot\bm{r}_{\text{A}\text{B}}}{|\bm{r}_{\text{A}\text{B}}|^3}} 
\right\}
\\
\nonumber 
&& +\frac{1}{c^2}\sum\limits_{\text{A}}
{GM_\text{A}} \frac{\dot{\bm{x}}_\text{s} - \dot{\bm{x}}_\text{A}}{ |\bm{r}_{\text{s}\text{A}}|^3}\,
\left\{ { 4\bm{r}_{\text{s}\text{A}}\cdot\dot{\bm{x}}_\text{s} -3\bm{r}_{\text{s}\text{A}}\cdot\dot{\bm{x}}_\text{A} }\right\} \\ 
\nonumber
&&- \frac{7}{ 2c^2}\sum\limits_{\text{A}}
{\frac{GM_\text{A}}{|\bm{r}_{\text{s}\text{A}}|}\sum\limits_{ \text{B}\ne \text{A}} { GM_\text{B}\, \frac{\bm{r}_{\text{A}\text{B}}} {|\bm{r}_{\text{A}\text{B}}|^3} } }
+ O(c^{-4}), 
\end{eqnarray}
where an overdot denotes the total time derivative with respect to $t=\text{TCB}$ (e.g., $\dot{\bm{x}}_s(t) = d\bm{x}_s(t)/dt$).

Finally, we highlight a relevant aspect of computing the acceleration due to a planet’s spherical harmonics. The multipole structure of the post-Newtonian gravitational field of a body can be adequately described only in the local reference system of that body. Therefore, the acceleration due to the spherical harmonics of a planet is usually computed in the local planetocentric system. 

In order to express this acceleration in the barycentric system, the method reported in \cite{moyer} is usually employed. Such a method consists of three steps: firstly, convert the barycentric position vector of the spacecraft to the planetocentric frame; then compute the acceleration due to the spherical harmonics of the planet in the local reference system; finally, convert this acceleration back from the planetocentric frame to the barycentric one. 

The formulas of coordinate transformations and conversion of accelerations employed in this method are based on the computations reported in \cite{hellings} and \cite{HRTW}, which in all scientific applications have been replaced by the IAU formulas. Moreover, we stress that in the above method, only the acceleration due to the spherical harmonics is converted. However, the formulas for the conversion of accelerations derived by the IAU models (Eq.~(\ref{eq: transf acc}) and its inverse) are non-linear. Therefore, separately transforming specific terms of the planetocentric equations of motion could lead to incorrect results. 

\subsection{A Mercury-centric reference system}
For the case of the BepiColombo radio science experiment, a local reference system for Mercury is required. Following \cite{Milani2010}, we introduce a Mercury-centric reference system with its time coordinate Mercury Coordinate Time (TCM) and its scaled version, the Mercury Time (TM). 

In the case of a Mercury-centric reference system, the above formulas for the space-time transformations are valid, provided that all the quantities are referred to Mercury. In particular, one possible choice for the constant $L_\text{M}$ is to set $L_\text{M} = L_\text{B}$ \cite{Milani2010}. This way, the mass parameters are the same in the scaled barycentric and Mercury-centric reference systems. Consequently, under this choice of the scaling, the constants derived from data analysis performed in the scaled Mercury local frame will be both TM-compatible and TDB-compatible. 

It is important to note that this choice of scaling is not the only option. Indeed, as stated previously, $L_\text{M}$ in principle could be chosen similarly to $L_\text{G}$ (so that the mean rate of TM matches the mean rate of the proper time of an observer situated on the rotating Mercury geoid), or it could be chosen such that $L_\text{M}=0$ (i.e., no scaled coordinates). Since TM-compatible accelerations in local coordinates are obtained by dividing the corresponding TCM-compatible accelerations by the factor $(1-L_\text{M})$ (see Section~\ref{sec: evaluation of the relativistic accelerations}), other choices of the scaling constant $L_\text{M}$ different from $L_\text{M} = L_\text{B}$ would only slightly change the values of the post-Newtonian accelerations, not affecting the general conclusions of this study. 

With the choice of scaling $L_\text{M} =L_\text{B}$, in order to compute the {TM} time (denoted in the following as $T^{**}$) at the Mercury center as a function of the TDB time, one solves the differential equation~(\ref{eq: ODE TDB TDX}) with $C_1=0$ and $C_2=1$, starting from the initial condition 
\begin{eqnarray}
\nonumber
    \Delta t^* && (\text{JD}_{\text{TDB}} = 2443144.5003725 \\
    \label{eq: initial condition}
    && - 6.55\times 10^{-5}/86400) = 6.55\times 10^{-5} \; \text{s}.
\end{eqnarray}

\begin{figure}[h!]
    \centering
  \includegraphics[width=\linewidth]{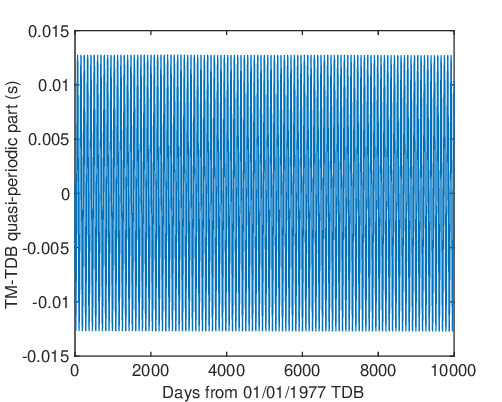}
\caption{ \label{fig: tdm-tdb periodic} Quasi-periodic part of the {TM}-TDB difference evaluated at the center of mass of Mercury from the 1st January 1997 TDB. This term has an amplitude of approximately 0.0127 s. The linear part of the transformation amounts to $-3.825\time10^{-8}$.}
\end{figure}

 The TM-TDB difference evaluated at the center of mass of Mercury comprises a linear term, which causes a drift between the two timescales, and a quasi-periodic term. The linear drift is approximately equal to $-3.825\times10^{-8}$, while the quasi-periodic term has an amplitude of approximately $0.0127$ s, as shown in Fig.~\ref{fig: tdm-tdb periodic}.

As stated previously, the TM-TDB transformation can be divided into the sum of the transformation at the center of Mercury and the location-dependent terms. In particular, for the case of BepiColombo, the MPO's altitude will vary approximately between 480 km and 1500 km, corresponding to $|\bm{r}_\text{M}| \lesssim 4000 \; \text{km}$, while the Mio's altitude will range between 590 km to 11639 km, i.e. $|\bm{r}_\text{M}| \lesssim 14100 \; \text{km}$. Therefore, location-dependent terms can be estimated in these two regions, and the results of such estimates are reported in Table~\ref{tab: estimates}. 

\begin{table}[h]
\caption{ \label{tab: estimates} Upper estimate of the location-dependent terms in the {TM}-TDB transformation (Eq.~(\ref{eq: time transformation})).}
\begin{ruledtabular}
    \begin{tabular}{c  c  c} 
         & $|\bm{r}_\text{M}| \lesssim 4000 \; \text{km}$ & $|\bm{r}_\text{M}| \lesssim 14100 \; \text{km}$\\ \colrule
         \rule{0pt}{2.4ex}
         $ v_\text{M}^i  r_\text{M}^i/c^2 $& $2.6 \times 10^{-6}$ s & $9.3 \times 10^{-6}$ s \\
         $ B^i(t) \, r_\text{M}^i/c^4$ & $3.0 \times 10^{-13}$ s  & $1.1 \times 10^{-12}$ s\\
         $ B^{ij}(t) \, r_\text{M}^i r_\text{M}^j/c^4$& $7.3 \times 10^{-18}$ s & $9.6 \times 10^{-17}$ s \\
         $ C(t, \bm{x})/c^4$ &$6.4 \times 10^{-23}$ s & $2.8 \times 10^{-21}$ s \\
    \end{tabular}
    \end{ruledtabular}
\end{table}

As shown in Table~\ref{tab: estimates}, the order of magnitude of the $O(c^{-4})$ terms is less than or equal to $10^{-12}$ seconds. Hence, those terms can be safely neglected. Therefore, for the purposes of the BepiColombo mission, the {TM}-TDB transformation can be written as 
\begin{equation}
\label{eq: time transformation bc}
    T^{**}=T^{**}_\text{center} -\frac{1}{c^2} v_\text{M}^i  r_\text{M}^i.
\end{equation}

\section{Numerical experiments}
\label{sec: numerical experiments}
\subsection{Evaluation of the relativistic accelerations}
\label{sec: evaluation of the relativistic accelerations}
As mentioned in Section~\ref{sec: introduction}, the equations of motion for a satellite in the Mercury-centric local frame include a local acceleration $\Phi_\text{M}^a$, describing Mercury’s gravitational field in the absence of an external world, a Mercury-third body coupling term $\Phi_\text{coup}^a$, and terms due to the influence of external bodies. The external accelerations consist of a ``gravito-electric'' term $\Phi_{\text{el}}^a$, which is independent of the velocity of the satellite, and a ``gravito-magnetic'' one $\Phi_\text{mg}^a$, which depends on the velocity of the satellite. The expressions for $\Phi_\text{M}^a$, $\Phi_\text{coup}^a$, $\Phi_{\text{el}}^a$ and $\Phi_\text{mg}^a$ are listed in Appendix~\ref{sec: eqs of motion} (see Eqs.~(\ref{eq:phiE})--(\ref{eq:phimg})). These equations contain the functions $W_\text{M}$, $W_\text{M}^a$, $W_\text{T}$, $W_\text{T}^a$, $Q_a$ and $C_b$, which come from the splitting of the local potentials. In particular, $W_\text{M}$ and $W_\text{M}^a$ represent the local internal gravitational potentials resulting from the gravitational attraction of Mercury, and can be expanded in terms of Mercury's post-Newtonian Blanchet and Damour's mass and spin multipole moments $\mathcal{M}^\text{M}_L$ and $\mathcal{S}^\text{M}_L$ \cite{DSX4}. The scalar potential $W_\text{M}$  can be equivalently expanded in terms of the spherical harmonic coefficients $\mathcal{C}^\text{M}_{lm}$ and $\mathcal{S}^\text{M}_{lm}$ (see Appendix~\ref{sec: phi_M}).
Moreover, $W_\text{T}$ and $W_\text{T}^a$ are external potentials representing tidal fields of other bodies of the system, the function $Q_a(T)$ expresses a deviation of the origin of the local frame from a geodesic of the external metric, and $C_a(T)$ accounts for inertial effects appearing in a kinematically non-rotating local reference system (i.e., Lense-Thirring, geodetic (or de Sitter) and Thomas precessions). Further details are reported in Appendix~\ref{sec: eqs of motion}. In particular, Appendix~\ref{sec: phi_M} deals with the description of the local acceleration $\Phi_\text{M}^a$, showing that, under suitable assumptions, the post-Newtonian part of $\Phi_\text{M}^a$ can be split as a sum of the Schwarzschild acceleration, which results from Mercury's mass monopole, the Lense-Thirring acceleration, arising from Mercury's spin vector, and the relativistic acceleration resulting from the quadrupole mass moments. In Appendix~\ref{sec: Qa} we present analytical and numerical estimates of the function ${{Q}}_a$ and its derivative $\dot{{Q}}_a= d{Q}_a/dT$.

As mentioned in Section~\ref{sec: introduction}, the BepiColombo mission comprises two spacecraft to be placed in different orbits around Mercury. The MPO will be placed in a polar orbit of 2.3h period, with altitude above the planet's surface ranging approximately between 480 km and 1500 km, while the altitude range for the Mio is expected to vary between 590 km and 11639 km. 

We implemented a quadruple-precision Fortran 90 code that evaluates the relativistic accelerations acting on the two Mercury orbiters in the Mercury-centric local reference system, using the ephemerides for the two spacecraft given by the ESA SPICE Kernels \cite{SPICE}. 
We summarize here the various assumptions adopted in our code to compute the post-Newtonian equations of motion. 
We approximated the external bodies (i.e., the Sun, the other planets, and the Moon) as point masses, while we considered Mercury with its extended structure. 

To compute the post-Newtonian part of $\Phi_\text{M}^a$ we neglected all time derivatives of the multipole moments $\dot{\mathcal{M}}^\text{M}_L$, $\ddot{\mathcal{M}}^\text{M}_L$, $\dots$, along with all mass multipole moments $\mathcal{M}^\text{M}_L$ for $l>2$, all spin moments $\mathcal{S}^\text{M}_L$ for $l>1$ and terms which are quadratic in Mercury's quadrupole moments $\mathcal{M}^\text{M}_{ab}$ or bilinear in $\mathcal{M}^\text{M}_{ab}$ and $\mathcal{S}^\text{M}_c$. Moreover, $\mathcal{M}^\text{M}_a=0$ as a consequence of the choice made in defining the local reference system (see Appendix~\ref{sec: phi_M}).  

To compute $\Phi_\text{coup}^a/c^2$, we decided to compute Mercury's gravitational potential and its derivative in Eq.~(\ref{eq:phicoup}) by taking into account only the mass monopole term. This choice has been driven by the smallness of the $\Phi_\text{coup}^a/c^2$ term, as it will be shown in the following. 

Moreover, to compute the accelerations $\Phi_\text{el}^a$, $\Phi_\text{mg}^a$ and $\Phi_\text{coup}^a$ according to Eqs.~(\ref{eq:phicoup})--(\ref{eq:phimg}) we neglected post-Newtonian terms containing ${{Q}}_a$ and $\dot{{Q}}_a$, since we estimated analytically that those terms are at most of the order of $10^{-22}$~m/s$^2$. Instead, we kept the Newtonian $Q_a$ in the expression for $\Phi_\text{el}^a$ (Eq.~(\ref{eq:phiel})). In the computation of $Q_a$, all multipole moments for $l\geq 3$ and all post-Newtonian terms have been discarded (see Appendix~\ref{sec: Qa}).

 Finally, we took the values for the spherical harmonic coefficients from the Mercury gravity solution HgM009 \cite{GENOVA2023115332}.

 Particular care must be taken when computing the acceleration due to external bodies, since spacetime transformations from local to global coordinates are required in order to retrieve their state vectors from the ephemerides. Starting from a {TM} time $T^{**}$ and {TM}-compatible state vector of the spacecraft $(\bm{X}_\text{s}^{**}, {\bm{V}}_\text{s})$, it is necessary to convert from local coordinates $(T^{**}, \bm{X}_\text{s}^{**})$ to global ones $(t^*, \bm{x}_\text{s}^{*})$, retrieve the state vectors of the planets, the Sun and the Moon at time $t^{*}$ from the ephemerides and then use Eqs.~(\ref{eq: W_T,a})--(\ref{eq: Cadot}) in order to compute $\Phi_\text{el}^a$, $\Phi_\text{mg}^a$ and $\Phi_\text{coup}^a$. 

 Note that the TM-compatible velocity coincides with the TCM-compatible one (that is, ${\bm{V}}_\text{s}^{**}= d{\bm{X}}_\text{s}^{**}/dT^{**}= d{\bm{X}}_\text{s}/dT={\bm{V}}_\text{s}$), while the TM-compatible acceleration ${\ddot{\bm{X}}}_\text{s}^{**}= d^2{\bm{X}}_\text{s}^{**}/dT^{**2}$ can be obtained by dividing the TCM-compatible acceleration ${\ddot{\bm{X}}}_\text{s}= d^2{\bm{X}}_\text{s}/dT^{2}$ by the factor $(1-L_\text{M})$ (i.e., ${\ddot{\bm{X}}}_\text{s}^{**}=  {\ddot{\bm{X}}}_\text{s}/(1-L_\text{M})$). Moreover, under the three scalings reported in Eqs.~(\ref{eq: TDX})--(\ref{eq: scaling mass parameter}), the equations of motion (Eq.~(\ref{eq:geodesic})) remain unchanged. This means that one can compute the TM-compatible acceleration of the spacecraft using Eq.~(\ref{eq:geodesic}) with TM-compatible quantities. 
 For simplicity of notation, in the following we will denote the TM-compatible accelerations $\Phi_\text{M}^{a\,**}$, $\Phi_\text{el}^{a\,**}$, $\Phi_\text{mg}^{a\,**}$, $\Phi_\text{coup}^{a\,**}$ simply as $\Phi_\text{M}^{a}$, $\Phi_\text{el}^{a}$, $\Phi_\text{mg}^{a}$, $\Phi_\text{coup}^{a}$.

\begin{figure}[h!]
    \centering
    \includegraphics[width=\linewidth]{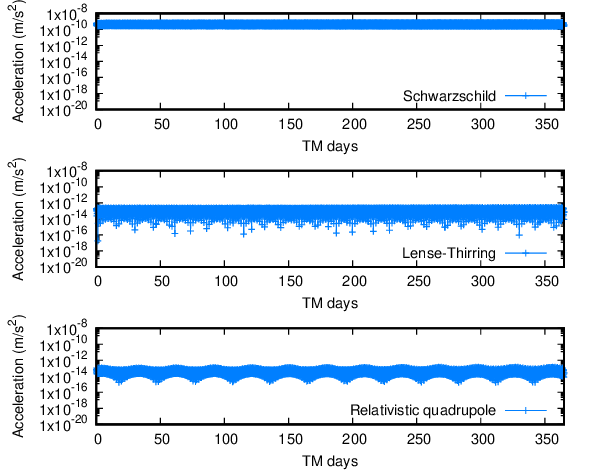}
    \caption{\label{fig: phiM_MPO} Norm of post-Newtonian terms in $\Phi_\text{M}^a$, evaluated on the MPO's trajectory.}
\end{figure}

 The magnitude of the post-Newtonian accelerations composing $\Phi_\text{M}^a$ is shown in Fig.~\ref{fig: phiM_MPO}, where the accelerations have been evaluated on the trajectory of the MPO during one year, starting from the 1st April 2027. The Schwarzschild term is the most important, reaching a magnitude of $\approx7.2\times10^{-10}$~m/s$^2$. The next term is the Lense-Thirring acceleration, measuring \mbox{$\approx2.8\times10^{-13}$~m/s$^2$}. The relativistic quadrupole term is $\approx1.4\times10^{-13}$~m/s$^2$.
 
\begin{figure}[h!]
    \centering
    \includegraphics[width=\linewidth]{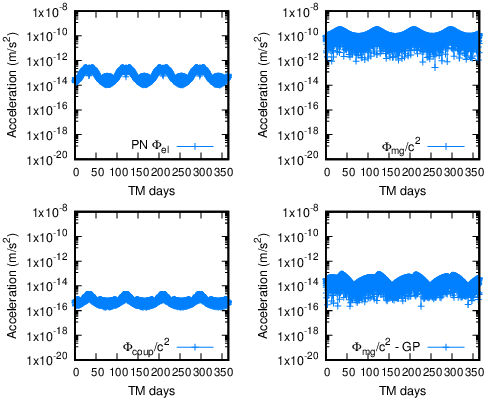}
    \caption{ \label{fig: phiext_MPO} Norm of post-Newtonian $\Phi_\text{el}^a$, $\Phi_\text{mg}^a/c^2$, $\Phi_\text{coup}^a/c^2$, and of the difference between $\Phi_\text{mg}^a/c^2$ and the geodetic precession, evaluated on the MPO's trajectory.}
\end{figure}

In Fig.~\ref{fig: phiext_MPO} we show the post-Newtonian accelerations due to external bodies, i.e., the post-Newtonian gravito-electric term $\Phi_\text{el}^a$ and the gravito-magnetic term $\Phi_\text{mg}^a/c^2$, as well as the Mercury-third body coupling term $\Phi_\text{coup}^a/c^2$, evaluated on the MPO orbit. These accelerations reach a maximum when the Mercury-Sun distance reaches its minimum. The most important acceleration is the gravito-magnetic term $\Phi_\text{mg}^a/c^2$, reaching a value of $\approx3.3\times10^{-10}$~m/s$^2$. This acceleration is mostly due to the geodetic precession caused by the Sun. Indeed, from Fig.~\ref{fig: phiext_MPO} one notices that the difference between $\Phi_\text{mg}^a/c^2$ and the geodetic precession of the Sun is less than $10^{-13}$~m/s$^2$. The next term is given by the post-Newtonian $\Phi_\text{el}^a$, reaching a maximum of $2.7\times 10^{-13}$~m/s$^2$. Finally, $\Phi_\text{coup}^a/c^2$ is the smallest term, with an order of magnitude of $10^{-15}$~m/s$^2$.

\begin{figure}[h!]
    \centering
    \includegraphics[width=\linewidth]{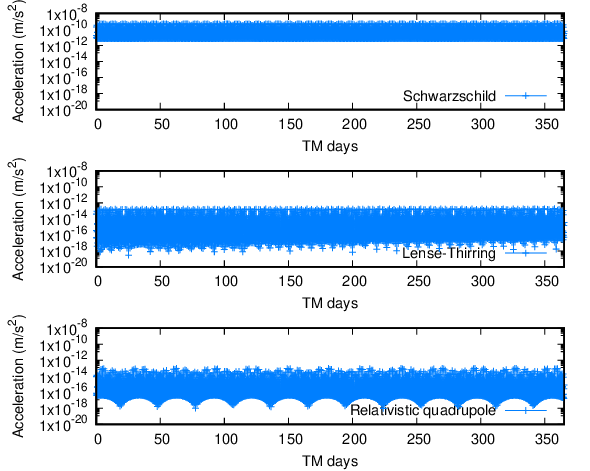}
    \caption{Norm of post-Newtonian terms in $\Phi_\text{M}^a$, evaluated on Mio's trajectory.}
    \label{fig: phiM_MMO}
\end{figure}

\begin{figure}[h!]
    \centering
    \includegraphics[width=\linewidth]{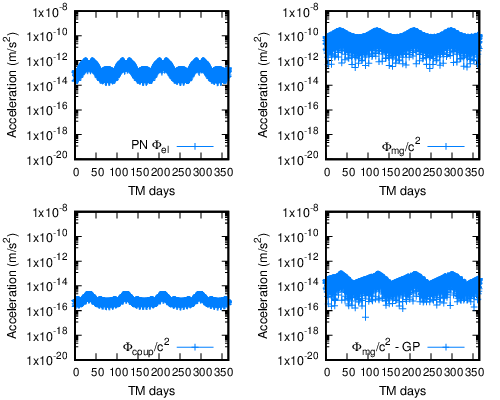}
    \caption{ \label{fig: phiext_MMO} Norm of post-Newtonian $\Phi_\text{el}^a$, $\Phi_\text{mg}^a/c^2$, $\Phi_\text{coup}^a/c^2$, and of the difference between $\Phi_\text{mg}^a/c^2$ and the geodetic precession, evaluated on Mio's trajectory.}
\end{figure}

 The relativistic accelerations have also been evaluated on Mio's trajectory for completeness, as shown in Fig.~\ref{fig: phiM_MMO} and Fig.~\ref{fig: phiext_MMO}.

We notice from Fig.~\ref{fig: phiM_MMO} that the maximum magnitudes of the Schwarzschild, Lense-Thirring, and relativistic quadrupole accelerations acting on Mio are very similar to the ones acting on the MPO spacecraft. This is because the two orbiters reach similar minimum altitudes on Mercury's surface. The three accelerations reach smaller values on Mio's trajectory, since Mio will orbit at a higher altitude with respect to the MPO. As shown in Fig.~\ref{fig: phiext_MMO}, the maximum values of the external and coupling terms on Mio's trajectory are similar to those reached on the MPO trajectory. The most significant difference is given by the post-Newtonian $\Phi_\text{el}^a$ term, whose norm on Mio's trajectory is $\approx 9.4 \times 10^{-13}$~m/s$^2$, which is more than 3 times the norm on the MPO's trajectory. 

The maximum values of the post-Newtonian acceleration along the MPO and Mio trajectories are reported in Table~\ref{tab: pn accelerations}.  
For comparison, the post-Newtonian accelerations shown in Table~\ref{tab: pn accelerations} are orders of magnitude smaller than the 
acceleration associated with the $C_{lm}^\text{M}(T)$ and $S_{lm}^\text{M}(T)$ multipoles in $W_{\text{M},a}$. 
By adopting the values for the spherical harmonic coefficients from the latest Mercury gravity solution HgM009 \cite{GENOVA2023115332}, we evaluated the accelerations associated with the zonal coefficients $C_{l0}^\text{M}$ along the MPO trajectory. For $l=2$, we obtained a maximum value of $\approx 2.3 \times 10^{-4}$~m/s$^2$. The accelerations due to higher-degree terms are also significant: for example, the $C^\text{M}_{20,0}$ and $C^\text{M}_{40,0}$ contributions have maximum magnitudes of $\approx 6.9 \times 10^{-7}$~m/s$^2$ and $1.2\times 10^{-8}$~m/s$^2$, respectively, still above the post-Newtonian Schwarzschild acceleration. 
A detailed assessment of which multipole coefficients are relevant for an accurate dynamical model of a Mercury orbiter is beyond the scope of this work. Nevertheless, Mercury's gravitational field should be modeled using a complete spherical harmonic expansion, including all relevant multipole coefficients $C_{lm}^\text{M}(T)$ and $S_{lm}^\text{M}(T)$.

\begin{table}[h]
\caption{\label{tab: pn accelerations} Maximum values of the post-Newtonian accelerations in the Mercury-centric reference system, evaluated along the MPO and Mio trajectories over a one-year period starting on 1 April 2027.}
\begin{ruledtabular}
    \begin{tabular}{c  c  c} 
    \multicolumn{1}{c}{Post-Newtonian} & \multicolumn{2}{c}{Max value (m/s$^2$)} \\
         acceleration & MPO  &  Mio  \\ \colrule
          \rule{0pt}{2.8ex}
         Schwarzschild &  $7.2 \times 10^{-10}$  & $5.2 \times 10^{-10}$   \\
         Lense-Thirring & $2.8 \times 10^{-13}$  &$1.9 \times 10^{-13}$  \\
         Relativistic quadrupole & $1.4 \times 10^{-13}$  & $1.0 \times 10^{-13}$  \\
    $\left[\bm{\Phi}_\text{el}\right]_\text{PN}$ & $2.7 \times 10^{-13}$   & $9.4 \times 10^{-13}$  \\
    \rule{0pt}{2.4ex}
    $\bm{\Phi}_\text{mg}/c^2$ & $3.3 \times 10^{-10}$  &  $2.7 \times 10^{-10}$  \\
    $\bm{\Phi}_\text{coup}/c^2$ &$2.0 \times 10^{-15}$ & $2.0 \times 10^{-15}$  \\
    \end{tabular}
    \end{ruledtabular}
\end{table}

\subsection{Comparison with the acceleration computed in the barycentric system}
Eq.~(\ref{eq: transf acc}) expresses the relation between the acceleration of the spacecraft $\ddot{\bm{x}}_\text{s}(t)=d^2\bm{x}_\text{s}(t)/dt^2$ in the barycentric frame and the acceleration $\ddot{\bm{X}}_\text{s}(T)=d^2\bm{X}_\text{s}(T)/dT^2$ in the local planetocentric frame. We can use this formula to validate our implementation of the Mercury-centric accelerations. This test can be made by comparing the acceleration computed in the barycentric system with the acceleration calculated in the planetocentric system, under the simplifying assumption that Mercury is considered a point mass and lacks an extended structure.  

Because of the choice $L_\text{M}=L_\text{B}$, Eq.~(\ref{eq: transf acc}), as well as Eqs.~(\ref{eq: transf pos spacecraft}) and (\ref{eq: transf velocity}), can be used with TDB-compatible quantities (on the right-hand side) and {TM}-compatible quantities (on the left-hand side) without making any changes to the formulas. Therefore, we can use Eq.~(\ref{eq: transf acc}) to compare the TDB-compatible acceleration $\ddot{\bm{x}}^*_\text{s}(t^*)=d^2\bm{x}^*_\text{s}(t^*)/dt^{*2}$ of the spacecraft in the barycentric reference system with the TM-compatible acceleration $\ddot{\bm{X}}^{**}_\text{s}(T^{**})=d^2\bm{X}^{**}_\text{s}(T^{**})/dT^{**2}$ in the local planetocentric reference system. 

In the following, for simplicity of notation, we will drop the superscripts and simply denote as $\ddot{\bm{x}}_\text{s}$ and ${\bm{A}}_\text{s}$ the TDB-compatible barycentric acceleration and the TM-compatible planetocentric acceleration, respectively. 

The TDB-compatible acceleration $\ddot{\bm{x}}_\text{s}$ in the barycentric reference system can be computed with Eq.~(\ref{EIH-eqm}), since the three scalings Eqs.~(\ref{eq: TDB})--(\ref{eq: scaling mass parameter TDB}) leave the equations of motion (Eq.~(\ref{EIH-eqm})) unchanged. The acceleration $\ddot{\bm{x}}_\text{s}$ can be substituted in Eq.~(\ref{eq: transf acc}) to get the TM-compatible converted acceleration $\hat{\bm{A}}_\text{s}$ from the barycentric system to the planetocentric one.

We expect the converted acceleration $\hat{\bm{A}}_\text{s}$ to be equivalent to the TM-compatible acceleration ${\bm{A}}_\text{s}$ computed from the geodetic equation (Eq.~(\ref{eq:geodesic})) up to $O(c^{-4})$ terms, i.e.:
\begin{equation}
\hat{\bm{A}}_\text{s}-\bm{A}_\text{s}=O(c^{-4}).
\end{equation}
The terms of order $O(c^{-4})$ appear here because we are using post-Newtonian expressions to compute the difference between the two accelerations, and post-post-Newtonian terms naturally appear during the algebraic manipulations.

The practical steps required to perform this test are the following: 
\begin{enumerate}
    \item at a TDB time $t^*$, the barycentric state vectors of all planets and the Sun are retrieved from the ephemerides, and the barycentric state vector of the orbiter is taken from BepiColombo's SPICE Kernels \cite{SPICE}. Note that all these quantities are TDB-compatible;
    \item the TDB-compatible acceleration $\ddot{\bm{x}}_\text{s}$ is computed using Eq.~(\ref{EIH-eqm}), with state vectors and mass parameters $\mu^{*}_\text{A}=GM_\text{A}^{*}$ of all the bodies of the solar system taken to be TDB-compatible;
    \item the TDB time $t^{*}$ is converted to the {TM} time $T^{**}$ at the location of the spacecraft with Eq.~(\ref{eq: time transformation bc});
    \item the TDB-compatible barycentric position vector of the orbiter and its velocity at time $t^{*}$ are converted to the TM-compatible Mercury-centric position and velocity vectors at time $T^{**}$, using Eq.~(\ref{eq: transf pos spacecraft}) and (\ref{eq: transf velocity});
    \item the TM-compatible acceleration of the spacecraft $\bm{A}_\text{s}$ is computed in the Mercury-centric system using Eq.~(\ref{eq:geodesic}) in the mass-monopole approximation;
    \item the TDB-compatible barycentric acceleration $\ddot{\bm{x}}_\text{s}$ is converted to the Mercury-centric system with Eq.~(\ref{eq: transf acc}). Note that the converted acceleration $\hat{\bm{A}}_\text{s}$ will be TM-compatible; 
    \item finally, the converted acceleration $\hat{\bm{A}}_\text{s}$ is compared with ${\bm{A}}_\text{s}$.  
\end{enumerate}

\begin{figure}[h!]
    \centering
    \includegraphics[width=\linewidth]{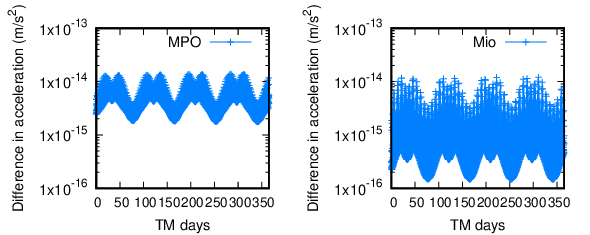}
    \caption{Norm of the difference $\hat{\bm{A}}_\text{s}- {\bm{A}}_\text{s}$ evaluated on the MPO and Mio's trajectories.}
    \label{fig: diff}
\end{figure}

The difference $\hat{\bm{A}}_\text{s}- {\bm{A}}_\text{s}$ computed on the MPO and Mio's trajectories during one year is shown in Fig.~\ref{fig: diff}. The difference reaches a maximum of $1.4\times10^{-14}$~m/s$^2$ on the MPO trajectory and $1.2\times10^{-14}$~m/s$^2$ on Mio's trajectory. To validate our implementation, we verified that the resulting difference contains only terms of post-post-Newtonian order. This was done by artificially increasing the value of $c$ by a factor of $10$, which led to a decrease by a factor of $10^{4}$ in the computed difference of accelerations, consistent with the expected behavior for the terms of order $O(c^{-4})$. It is important to note that these post-post-Newtonian terms have no physical significance, as they do not originate from the post-post-Newtonian equations of motion but rather result purely from algebraic manipulations of the post-Newtonian equations.  Moreover, the physical post-post-Newtonian terms in the equations of motion for both the barycentric and planetocentric reference systems remain unknown, as these equations have not yet been explicitly derived to post-post-Newtonian order. As a result, it is not possible to have full control over the $O(c^{-4})$ terms.

Note that a similar test was performed in \cite{Morley}, in order to check for the compatibility of the trajectories obtained by integrating the equations of motion in the BCRS and in the local Mercury-centric system. The authors found a maximum value of the difference of accelerations $\hat{\bm{A}}_\text{s}- {\bm{A}}_\text{s}$ computed along the MPO trajectory equal to $\approx5\times10^{-13}$~m/s$^2$. However, their approach differs from ours because they considered a simpler version of the Mercury-centric equations of motion, comprising only the Schwarzschild acceleration and the geodetic precession caused by the Sun. Moreover, they used a slightly different formula for the conversion of accelerations, in which they discarded all the terms containing the first and second derivatives of the acceleration of Mercury, and the second derivative of the gravitational potential of the external bodies. 

\subsection{Integration of the equations of motion}
Analysis of MORE radio tracking data collected during the initial cruise phase of the BepiColombo mission showed that range measurements achieved 1-centimeter-level accuracy \cite{cappuccio}. To assess the contribution of the post-Newtonian accelerations to the trajectory of the orbiter, we integrated the equations of motion of the MPO in the local Mercury-centric system. The orbit obtained by numerical integration of the full equations of motion in the local frame (Eq.~(\ref{eq:geodesic})) is used as a reference trajectory. We compare such a reference trajectory with the orbits obtained by integrating the equations of motion without specific post-Newtonian terms. 

The current nominal configuration of the MORE experiment assumes that tracking will be carried out from two ground antennas: ESA’s DSA 3,
located in Malarg\"{u}e (Argentina) and NASA’s DSS 25, situated in Goldstone (California). The final tracking schedule will depend on factors such as spacecraft visibility, elevation above the local horizon, and operational constraints \cite{iess}. To perform the orbit determination of the spacecraft, a multi-arc strategy will be adopted \cite{Milani_Gronchi_2009}. The multi-arc strategy allows the solution to absorb possible uncertainties in the dynamical model (e.g., uncertainties in the modeling of non-gravitational perturbations, or dynamical maneuvers). According to this method, the trajectory will be divided into arcs of approximately 24 hours, each with independent initial conditions and including one daily tracking session from both stations, with data gaps expected between consecutive tracking passes.

\begin{figure}[h!]
    \centering
    \includegraphics[width=\linewidth]{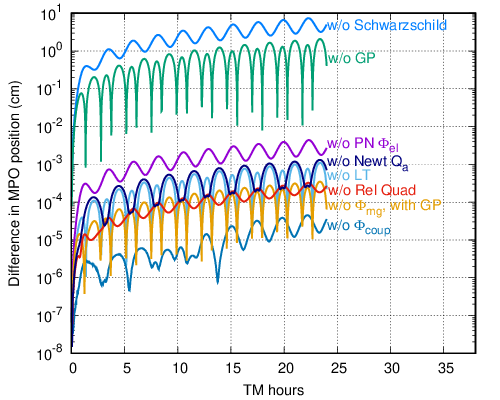}
    \caption{Difference in MPO position between the MPO reference trajectory (obtained by integrating the full equations of motion in the Mercury-centric frame) and a trajectory obtained from the equations of motion where one specific acceleration has been removed.}
    \label{fig: traj diff}
\end{figure}

\begin{figure}[h!]
    \centering
    \includegraphics[width=\linewidth]{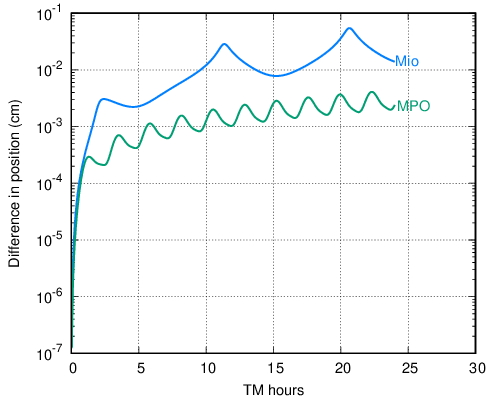}
    \caption{Difference in position between the reference trajectory for each of the two spacecraft (obtained by integrating the full equations of motion in the Mercury-centric frame) and the trajectory obtained with the truncated model, which includes the Schwarzschild acceleration and the geodetic precession of the Sun as the only post-Newtonian perturbations.}
    \label{fig: truncated model}
\end{figure}

 We performed such a comparison by integrating the several MPO trajectories on the 5th of May 2027, when the Mercury-Sun distance reaches its minimum and the external terms of the Mercury-centric acceleration reach their maximum magnitude. We considered a propagation arc of 24 hours, and we performed the integrations of the equations of motion using quadruple precision. The results are shown in Fig.~\ref{fig: traj diff}. The difference between the reference trajectory and the trajectory obtained without including the Schwarzschild term in the equations of motion is the most important, equal to $\approx8$~cm after 24 hours of numerical integration. The contribution of the geodetic precession of the Sun is also significant, with a difference of $\approx2$~cm after 24 hours. Finally, as shown in Fig.~\ref{fig: traj diff}, the other post-Newtonian terms are not so significant and can be safely neglected for the purposes of the radio science experiment onboard the BepiColombo mission. Indeed, the difference between the reference trajectory and the trajectory obtained including the Schwarzschild acceleration and the geodetic precession of the Sun as the only post-Newtonian perturbations is less than $10^{-2}$~cm (see Fig.~\ref{fig: truncated model}). 

For completeness, we repeated the same comparison for Mio. We found similar results, with the Schwarzschild term and the geodetic precession of the Sun being the most important accelerations. In particular, due to Mio's higher altitude, the geodetic precession is the most relevant perturbation, and the difference between Mio's reference trajectory and the trajectory obtained without including the geodetic precession is $\approx8$~cm after 24 hours of numerical integration. The contribution of the Schwarzschild term is $\approx3$~cm after 24 hours. Similarly to the MPO, also for Mio the other post-Newtonian perturbations are negligible, as the difference between Mio's reference trajectory and the trajectory obtained including the Schwarzschild acceleration and the geodetic precession of the Sun as the only post-Newtonian accelerations is less than $10^{-1}$~cm after 24 hours (see Fig.~\ref{fig: truncated model}). This difference is slightly higher than for the MPO and is mainly due to the post-Newtonian term of $\bm{\Phi}_\text{el}$. Indeed, the difference between the reference trajectory and the trajectory obtained without including the post-Newtonian $\bm{\Phi}_\text{el}$ term in the equations of motion is $\approx6\times10^{-2}$~cm, while for the MPO it was equal to $4\times10^{-3}$~cm. 

Finally, we remark that for practical applications, the relativistic modeling of spacecraft motion presented in this work should be complemented with a relativistic description of observables, including the modeling of light propagation and the process of observation itself. However, this description is beyond the scope of this work.

\section{Conclusions}
\label{sec: conclusions}
In this study, we investigated the modeling of spacecraft motion in a Mercury-centric local frame within the framework of general relativity. Indeed, the acceleration due to the multipole expansion of Mercury's post-Newtonian gravitational field can be appropriately described only in Mercury's local coordinates. Furthermore, as explained in Section~\ref{sec: transformations of accelerations}, the transformation of accelerations from the local to the global reference systems and vice versa is non-linear. Hence, from the point of view of the theory of astronomical reference systems, it is not correct to separately compute the acceleration due to Mercury's spherical harmonics in local coordinates, then convert it to the barycentric frame and sum it to the post-Newtonian acceleration computed in the barycentric frame.   

We implemented the full set of post-Newtonian equations of motion in the Mercury-centric frame in a dedicated Fortran 90 code. We presented analytical and numerical estimates of the terms composing the equations of motion along the trajectories of the BepiColombo MPO and Mio spacecraft that we retrieved from the SPICE kernels. We verified the correctness of our implementation by checking that our results are consistent with the equations of motion in the barycentric frame to the $O(c^{-4})$ level. 

We computed a reference orbit for the MPO spacecraft by integrating the full post-Newtonian equations of motion in the Mercury-centric frame under realistic mission assumptions.
To assess the contribution of each post-Newtonian term of the equations of motion, we compared the reference orbit with the trajectory obtained by integrating the Mercury-centric equations of motion without each post-Newtonian perturbation.   
We showed that the magnitude of the majority of post-Newtonian perturbations is too small to be relevant for the MORE experiment, and we concluded that the inclusion in the model of the Schwarzschild term and the geodetic precession is sufficient. 

However, our results show that already for a medium-flying satellite as Mio, the post-Newtonian term of the gravito-electric acceleration $\Phi_{\text{el}}$ reaches a value of \mbox{$\approx10^{-12}$~m/s$^2$}. While still small, this suggests that in future missions to Mercury requiring a highly accurate modeling, the inclusion of more post-Newtonian perturbations in the local equations of motion should be considered.

Finally, the practical approach proposed in this study may also prove valuable for the high-accuracy modeling of orbiters around other planets, as well as for spacecraft performing flyby maneuvers, where modeling in the barycentric frame is typically adopted. 
\begin{acknowledgments}
M.F. acknowledges the support of the Italian Space Agency (ASI) through ASI/CRAS agreement No. 2022-16-HH.0. G.R.M. is funded by the European Union's Horizon 2020 research and innovation program under the Marie Skłodowska-Curie grant agreement No. 101072454 (MWGaiaDN).
\end{acknowledgments}

\appendix
\section{Equations of motion in the local frame}
\label{sec: eqs of motion}

The equations of motion for a test particle in the Mercury-centric local frame can be derived from the geodesic equation. In the first post-Newtonian approximation of general relativity, the equations of motion for a satellite in closed form read \cite{relcelmechPPN}
\begin{equation}
\label{eq:geodesic}
    \Ddot{X}_s^a = \Phi_\text{M}^a + \Phi_{\text{el}}^a+ \frac{1}{c^2}(\Phi_\text{coup}^a+\Phi_\text{mg}^a) + O(c^{-4})
\end{equation}
where $\Phi_\text{M}^a$ contains the internal terms, describing Mercury's gravitational field in the absence of an external world; $\Phi_\text{coup}^a$ is the Mercury-third body coupling term; $\Phi_{\text{el}}^a$ and $\Phi_\text{mg}^a$ are terms due to the influence of external bodies. The former (the so-called ``gravito-electric" term) is independent of the velocity of the satellite, while the latter (the ``gravito-magnetic" term) depends on the velocity of the satellite.
The expressions for $\Phi_\text{M}^a$, $\Phi_\text{coup}^a$, $\Phi_{\text{el}}^a$ and $\Phi_\text{mg}^a$ read (see Eqs.~(8.3)--(8.6) of \cite{relcelmechPPN})
\begin{eqnarray}
    \nonumber
     \Phi_\text{M}^a & = & W_{\text{M},a} + \frac{1}{c^2} \left( 4 W_{\text{M},T}^a -4W_{\text{M}}W_{\text{M},a} -3W_{\text{M},T}\dot{X}_s^a \right. \\ 
        \nonumber
         &&  -4W_{\text{M},b}\dot{X}_s^b\dot{X}_s^a + W_{\text{M},a}\dot{X}_s^b\dot{X}_s^b\\ 
         \label{eq:phiE}
         && \left. +4\dot{X}_s^b(W_{\text{M},b}^a-W_{\text{M},a}^b) \right)+O(c^{-4}),  \\ 
        \nonumber
         \Phi_{\text{coup}}^a  &=&  -4\left(W_{\text{M}}(Q_a + W_{\text{T},a}) + W_{\text{M},a}(Q_b X_s^b+W_{\text{T}}) \right) \\
         \label{eq:phicoup}
         && + O(c^{-2}), \\
         \nonumber
         \Phi_{\text{el}}^a  &= &Q_a + W_{\text{T},a} \\
         \nonumber
         &&+ \frac{1}{c^2} \left( 4W_{\text{T},T}^a - 4 (Q_b X_s^b +W_{\text{T}})(Q_a+W_{\text{T},a}) \right. \\
        \label{eq:phiel}
        && \left. +2\epsilon_{abc}\dot{C}_bX_s^c\right) +O(c^{-4}), \\
        \nonumber
        \Phi_{\text{mg}}^a &=& -3(W_{\text{T},T}+ \dot{Q}_bX_s^b)\dot{X}_s^a - 4(W_{\text{T},b}+Q_b)\dot{X}_s^b\dot{X}_s^a \\
        \nonumber
        && +(W_{\text{T},a}+Q_a)\dot{X}_s^b\dot{X}_s^b + 8W_{\text{T}}^{[a,b]}\dot{X}_s^b  \\
        \label{eq:phimg}
        & &  +4\epsilon_{abc}C_b\dot{X}_s^c + O(c^{-2}). 
\end{eqnarray}
In Eqs.~(\ref{eq:phiE})--(\ref{eq:phimg}) the potentials $W_\text{M}$, $W_\text{T}$ and their derivatives should be evaluated at $(T,\bm{X})=(T, \bm{X}_\text{s}(T))$, where $\bm{X}_\text{s}$ are the spatial coordinates of the spacecraft. An overdot here denotes the total time derivative with respect to $T=\text{TCM}$ (e.g. $\dot{X}_s^a=d{X}_s^a/dT$).
The functions $W_\text{M}$, $W_\text{M}^a$, $W_\text{T}$, $W_\text{T}^a$, $Q_a$ and $C_b$ come from the splitting of the local potentials: 
\begin{eqnarray*}
     W(T, \bm{X}) &=& W_\text{M}(T, \bm{X})+Q_a(T)X^a+ W_\text{T}(T, \bm{X}),\\
     W^a(T, \bm{X}) &=& W_\text{M}^a(T, \bm{X})+\frac{1}{2}\epsilon_{abc}C_b(T)X^c+ W_\text{T}^a(T, \bm{X}).
\end{eqnarray*}

Particularly, $W_\text{M}$ and $W_\text{M}^a$ are the local internal gravitational potentials which result from the gravitational attraction of Mercury, $W_\text{T}$ and $W_\text{T}^a$ are external potentials describing tidal fields of other bodies of the system, the function $Q_a(T)$ is related to a deviation of the origin of the local frame from a geodesic of the external metric, and $C_a(T)$ describes inertial effects appearing in a kinematically non-rotating local reference system (i.e., Lense-Thirring, geodetic (or de Sitter) and Thomas precessions). 

The function $C_a(T)$ solves the equation (see Eq.~(4.12) in \cite{relcelmechPPN})
\begin{eqnarray}
\nonumber
    c^2 R_i^a \frac{d}{dT}{R}_j^a &=& -2\epsilon_{ijk}R_k^a C_a -4 \overline{w}^{[i,j]}(\bm{x}_\text{M}) + 3 v_\text{M}^{[i}\overline{w}_{,j]}(\bm{x}_\text{M}) \\
    \label{eq: Ca}
    && +v_\text{M}^{[i} R^a_{j]}Q_a + O(c^{-2}),
\end{eqnarray}
which, in the case of a kinematically non-rotating local system (i.e. $R_i^a=\delta_i^a$), can be more conveniently rewritten as:
\begin{eqnarray}
    \nonumber
    \epsilon_{ijk}{C_a}\delta^a_j&=&\overline{w}^i_{,k}(\bm{x}_\text{M})-\overline{w}^k_{,i}(\bm{x}_\text{M})\\
    \label{eq: Ca2}
    && +\frac{3}{4}\left(v_\text{M} ^k \overline{w}_{,i}(\bm{x}_\text{M})-v_\text{M} ^i\overline{w}_{,k}(\bm{x}_\text{M})\right)\\
    \nonumber
    &&+\frac{1}{4}\left(v_\text{M} ^k Q_a\delta_i^a - v_\text{M} ^i Q_a\delta^a_k\right) + O(c^{-2}).
\end{eqnarray}
Note that Eq.~(\ref{eq: Ca2}) contains the well-known expressions for the Lense-Thirring, geodetic (or de Sitter), and Thomas precessions.

\begin{widetext}
The tidal potentials $W_\text{T}$, $W_\text{T}^a$ admit an expression in terms of the external global potentials $\overline{w}$, $\overline{w}^i$. Such expressions are shown in Eqs.~(4.31) and (4.32) of \cite{relcelmechPPN} and read  
\begin{eqnarray}
    \nonumber
    W_\text{T}(T, \bm{X}) &=& \overline{w}(t,\bm{x}) - \overline{w}(\bm{x}_\text{M}) - \overline{w}_{,j}(\bm{x}_\text{M}) r^j_\text{M} + \frac{1}{c^2} \left\{-4 v^i_\text{M} \left[ \overline{w}^i(t, \bm{x})  - \overline{w}^i(\bm{x}_\text{M}) - \overline{w}^i_{,j}(\bm{x}_\text{M}) r^j_\text{M} \right]  \right.  \\
    \label{eq: W_T}
    &&+ 2 |\bm{v}_\text{M}|^2 \left[ \overline{w}(t, \bm{x})  - \overline{w}(\bm{x}_\text{M}) - \overline{w}_{,j}(\bm{x}_\text{M}) r^j_\text{M} \right]  + 2 \dot{\overline{w}}^{i}_{,j}(\bm{x}_\text{M}) r^i_\text{M} r^j_\text{M} +\frac{1}{2} \ddot{\overline{w}}(\bm{x}_\text{M}) |\bm{r}_\text{M}|^2  - \frac{3}{2} \left( a_\text{M}^i r^i_\text{M} \right)^2    \\
    \nonumber
    && \left. -3 Q_a \delta_k^a r_\text{M}^k  a^i_\text{M} r^i_\text{M} -  v^i_\text{M} r^i_\text{M}\dot{\overline{w}}_{,j}(\bm{x}_\text{M}) r^j_\text{M}   + \frac{1}{2}  |\bm{r}_\text{M} |^2 \delta^a_i a^i_\text{M} Q_a -\frac{1}{10} |\bm{r}_\text{M} |^2\ddot{a}_{\text{M}}^i r_\text{M}^i \right\} + {O}(c^{-4}),   \\
   \nonumber
    W^a_\text{T}(T, \bm{X}) & =& \delta^a_i \left\{ \overline{w}^i(t,\bm{x}) - \overline{w}^i(\bm{x}_\text{M}) - \overline{w}^i_{,j}(\bm{x}_\text{M})r_\text{M}^j- v^i_\text{M} \left(\overline{w}(t,\bm{x}) - \overline{w}(\bm{x}_\text{M}) - \overline{w}_{,j}(\bm{x}_\text{M})r_\text{M}^j\right)\right. \\ 
    \label{eq: W_T^a}
     && \left.  + \frac{3}{10} r^i_\text{M}\left(\dot{a}_\text{M}^k r^k_\text{M}\right) -\frac{1}{10} \dot{a}_\text{M}^i |\bm{r}_\text{M}|^2\right\}
     + O(c^{-2}).
\end{eqnarray}
Here and in the remaining formulas of this Section, a dot over functions of time $t=\text{TCB}$ designates the total time $d/dt$ with respect to time $t$, while a dot over functions of time $T=\text{TCM}$ denotes the total time derivative $d/dT$ with respect to time $T$. Note that in the post-Newtonian terms, there is no need to distinguish between $d/dt$ and $d/dT$, as the difference between these two timescales is of order $O(c^{-2})$. 

Briefly, Eqs.~(\ref{eq: W_T}) and (\ref{eq: W_T^a}) can be rewritten as $W_\text{T}(T,  \bm{X})= \widetilde{W}_\text{T}(t,\bm{x})$ and $W_\text{T}^a(T,  \bm{X})= \widetilde{W}^a_\text{T}(t,\bm{x})$, where the coordinates $(t, \bm{x})$ in the right-hand side are related to the coordinates $(T, \bm{X})$ by the 4-dimensional coordinate transformation.
\end{widetext}

 From such expressions, one then computes the derivatives $W_{\text{T}, a}$, $W_{\text{T}, T}$, $W_{\text{T},b}^a$,$W_{\text{T}, T}^a$ by calculating the Jacobian of the inverse transformation of coordinates $\frac{\partial x^\mu}{\partial X^\alpha}$ and by using the chain rule, i.e.  
 \begin{eqnarray*}
    W_{\text{T}, \alpha} &  =&  \widetilde{W}_{\text{T}, \mu} \frac{\partial x^\mu}{\partial X^\alpha}=\widetilde{W}_{\text{T},t}\frac{\partial t}{\partial X^\alpha}+\widetilde{W}_{\text{T},i} \frac{\partial x^i}{\partial X^\alpha},\\
    W^a_{\text{T}, \alpha} & =& \widetilde{W}^a_{\text{T},\mu} \frac{\partial x^\mu}{\partial X^\alpha}= \widetilde{W}^a_{\text{T},t} \frac{\partial t}{\partial X^\alpha}+ \widetilde{W}^a_{\text{T},i} \frac{\partial x^i}{\partial X^\alpha}. \\
 \end{eqnarray*}

Using the relations
\begin{eqnarray*}
    \frac{\partial t}{\partial T} &=& 1+O(c^{-2}),\\
    \frac{\partial x^i}{\partial T} &=&v_\text{M}^i + O(c^{-2}),\\
    \frac{\partial t}{\partial X^a} &= &\frac{\delta_i^a}{c^2} v_\text{M}^i + O(c^{-4}),\\
    \frac{\partial x^i}{\partial X^a} &=& \delta^i_a + \frac{1}{c^2}\left( \frac{1}{2} v_\text{M}^i v_\text{M}^k \delta^k_a - \delta^i_a\overline{w}(\bm{x}_\text{M}) + a_\text{M}^i r_\text{M}^k\delta^k_a\right. \\
    && \left.- a_\text{M}^k r_\text{M}^i\delta^k_a - a_\text{M}^k r_\text{M}^k\delta^i_a\right) + O(c^{-4}),\\
\end{eqnarray*}
\begin{widetext}
one obtains
\begin{eqnarray}
    \nonumber
   W_{\text{T},a} &=& \delta^i_a \left\{ \overline{w}_{,i}(t, \bm{x}) - \overline{w}_{,i}(\bm{x}_\text{M}) + \frac{1}{c^2}\left[\left( \overline{w}_{,t}(t, \bm{x}) - \dot{\overline{w}}(\bm{x}_\text{M}) - \dot{\overline{w}}_{,j}(\bm{x}_\text{M})r_\text{M}^j+ \overline{w}_{,j}(\bm{x}_\text{M})v_\text{M}^j \right)v_\text{M}^i  \right. \right. \\
     \nonumber
    &&  + \left( \overline{w}_{,j}(t, \bm{x}) -  \overline{w}_{,j}(\bm{x}_\text{M})\right)\left(\frac{1}{2}v_\text{M}^jv_\text{M}^i + r_\text{M}^i a_\text{M}^j - r_\text{M}^j a_\text{M}^i \right) + \left(\overline{w}_{,i}(t, \bm{x}) - \overline{w}_{,i}(\bm{x}_\text{M})\right) \left( 2 |\bm{v}_\text{M}|^2- \overline{w}(\bm{x}_\text{M})  - a_\text{M}^k r_\text{M}^k\right)  \\
    \label{eq: W_T,a}
   &&   - 4v_\text{M}^k \left(  \overline{w}^k_{,i}(t, \bm{x}) -  \overline{w}^k_{,i}(\bm{x}_\text{M}) \right)  + 2 \dot{\overline{w}}^k_{,i}(\bm{x}_\text{M})r_\text{M}^k + 2 \dot{\overline{w}}^i_{,k}(\bm{x}_\text{M})r_\text{M}^k + \ddot{\overline{w}}(\bm{x}_\text{M})r_\text{M}^i - 3a_\text{M}^kr_\text{M}^k a_\text{M}^i  - 3Q_b\delta_i^b a_\text{M}^kr_\text{M}^k\\
   \nonumber
    && \left. \left.  -3 Q_b r_\text{M}^k \delta_k^b a_\text{M}^i - v_\text{M}^i \dot{\overline{w}}_{,k}(\bm{x}_\text{M})r_\text{M}^k -\dot{\overline{w}}_{,i}(\bm{x}_\text{M})v_\text{M}^kr_\text{M}^k  + r_\text{M}^i a_\text{M}^k Q_b \delta_k^b -\frac{1}{10}\ddot{a}_\text{M}^i |\bm{r}_\text{M}|^2 -\frac{1}{5}\ddot{a}_\text{M}^k r_\text{M}^k r_\text{M}^i\right]\right\} + O(c^{-4}),  \\
    \label{eq: W_T,T}
     W_{\text{T}, T} & =& \overline{w}_{,t}(t, \bm{x}) - \dot{\overline{w}}(\bm{x}_\text{M}) - \dot{\overline{w}}_{,j}(\bm{x}_\text{M})r_\text{M}^j+\overline{w}_{,j}(\bm{x}_\text{M})v_\text{M}^j +  v_\text{M}^j\left( \overline{w}_{,j}(t, \bm{x}) -\overline{w}_{,j}(\bm{x}_\text{M})\right) + O(c^{-2}),\\
     \label{eq: W_T^a,b}
    W^a_{\text{T}, b} \! & =& \!  \delta^a_i\delta^j_b \! \left\{\overline{w}^i_{,j}(t, \bm{x}) - \overline{w}^i_{,j}(\bm{x}_\text{M}) - v_\text{M}^i \left( \overline{w}_{,j}(t, \bm{x}) - \overline{w}_{,j}(\bm{x}_\text{M})\right) +\frac{3}{10}\dot{a}_\text{M}^k r_\text{M}^k \delta_j^i +\frac{3}{10}\dot{a}_\text{M}^j r_\text{M}^i -\frac{1}{5}\dot{a}_\text{M}^ir_\text{M}^j\right\}\! +O(c^{-2}),\\
    \nonumber
      W^a_{\text{T}, T} & = &\delta^a_i \left\{ \overline{w}^i_{,t}(t, \bm{x})-\dot{\overline{w}}^i(\bm{x}_\text{M})- \dot{\overline{w}}^i_{,j}(\bm{x}_\text{M})r_\text{M}^j+\overline{w}^i_{,j}(t,\bm{x})v_\text{M}^j   -a_\text{M}^i \left(\overline{w}(t, \bm{x}) -  \overline{w}(\bm{x}_\text{M}) - \overline{w}_{,j}(\bm{x}_\text{M})r_\text{M}^j \right) \right. \\ 
       \label{eq: W_T^a,T}
      && \left. + \frac{3}{10}r_\text{M}^i \ddot{a}_\text{M}^kr_\text{M}^k -\frac{1}{10} \ddot{a}_\text{M}^i |\bm{r}_\text{M}|^2  
        -v_\text{M}^i \left( \overline{w}_{,t}(t, \bm{x}) - \dot{\overline{w}}(\bm{x}_\text{M}) -\dot{\overline{w}}_{,j}(\bm{x}_\text{M})r_\text{M}^j + \overline{w}_{,j}(t, \bm{x})v_\text{M}^j\right) \right\} + O(c^{-2}). 
\end{eqnarray}

Analogously, from Eq.~(\ref{eq: Ca2}) one can compute the time derivative $\dot{C}_a(T)=d{C}_a(T) /dT$ as 
\begin{eqnarray}
   \nonumber
     \epsilon_{ijk}{\dot{C}_a}\delta^a_j&=&\dot{\overline{w}}^i_{,k}(\bm{x}_\text{M})-\dot{\overline{w}}^k_{,i}(\bm{x}_\text{M}) +\frac{3}{4}\left(a_\text{M}^k \overline{w}_{,i}(\bm{x}_\text{M}) + v_\text{M}^k \dot{\overline{w}}_{,i}(\bm{x}_\text{M}) - a_\text{M}^i \overline{w}_{,k}(\bm{x}_\text{M}) - v_\text{M}^i \dot{\overline{w}}_{,k}(\bm{x}_\text{M})\right) \\ 
     \label{eq: Cadot}
      && + \frac{1}{4}\left(a_\text{M} ^k Q_a\delta_i^a + v_\text{M} ^k \dot{Q}_a\delta_i^a- a_\text{M} ^i Q_a\delta^a_k - v_\text{M} ^i \dot{Q}_a\delta^a_k\right) + O(c^{-2}).
\end{eqnarray}
\end{widetext}

\subsection{Explicit expressions for $\Phi^a_\text{M}$}
\label{sec: phi_M}
The internal local gravitational potentials $W_\text{M}$ and $W_\text{M}^a$ admit an expansion in terms of Mercury's post-Newtonian Blanchet and Damour's mass and spin multipole moments $\mathcal{M}^\text{M}_L$ and $\mathcal{S}^\text{M}_L$ \cite{DSX4}, as 
\begin{eqnarray} 
\nonumber
 W_\text{M}(T, \bm{X}) & =& G \sum_{l\geq 0} \frac{(-1)^l}{l!}  \left[\mathcal{M}^\text{M}_L\partial_L\frac{1}{|\bm{X}|} \right.\\
\label{eq: W_M3}
&& \left. + \frac{1}{2c^2}\ddot{\mathcal{M}}^\text{M}_L\partial_L|\bm{X}|\right]+ O(c^{-4}), \\
\nonumber
W_\text{M}^a(T, \bm{X}) & =& -G \sum_{l\geq 1} \frac{(-1)^l}{l!}  \left[\dot{\mathcal{M}}^\text{M}_{aL-1}\partial_{L-1}\frac{1}{|\bm{X}|} \right. \\ 
\label{eq: W_Ma}
&&  \left. + \frac{l}{l+1}\epsilon_{abc}\mathcal{S}^\text{M}_{cL-1}\partial_{bL-1}\frac{1}{|\bm{X}|}\right] \\
\nonumber 
&& + O(c^{-2}),
\end{eqnarray}
where $L$ denotes the spatial multi-index $a_1a_2 \dots a_l$, and $\partial_L$ denotes the partial derivative of order $l$ i.e.,
$\partial_L=\frac{\partial^l}{ \partial X^{a_1}\dots \partial X^{a_l}}$. Here and in the following, a dot over Mercury's mass multipole moments $\mathcal{M}^\text{M}_L$ denotes the total time derivative with respect to time $T=\text{TCM}$ (e.g., $\dot{\mathcal{M}}^\text{M}_L=d\mathcal{M}^\text{M}_L/dT$, $\ddot{\mathcal{M}}^\text{M}_L=d^2\mathcal{M}^\text{M}_L/dT^2$). 

Eq.~(\ref{eq: W_M3}) is equivalent to the spherical harmonic expansion \cite{Soffel2003}
\begin{widetext}
\begin{equation}
\label{eq: WM mathcal spher}
    W_\text{M}(T, \bm{X})   = \frac{G M_\text{M}}{|\bm{X}|} \left[ 1 + \sum_{l=2}^{\infty} \sum_{m=0}^{l} \left( \frac{R_\text{M}}{|\bm{X}|} \right)^l P_{lm}(\cos\theta) \left( \mathcal{C}^\text{M}_{lm}(T,|\bm{X}| ) \cos m\phi + \mathcal{S}^\text{M}_{lm}(T,|\bm{X}|) \sin m\phi \right) \right]  + O(c^{-4}),
\end{equation}
\end{widetext}
where $M_\text{M}$ and $R_\text{M}$ indicate Mercury's mass and radius, respectively; $P_{lm}$ are the Legendre associated functions of degree $l$ and order $m$; ($|\bm{X}|$, $\theta$, $\phi$) is the representation in spherical coordinates (distance, latitude, longitude) of the position of the spacecraft in the local frame, and $\mathcal{C}^\text{M}_{lm}$ and $\mathcal{S}^\text{M}_{lm}$ are given by
\begin{equation}
\label{eq: mathcal Clm}
    \mathcal{C}^\text{M}_{lm}(T, |\bm{X}|)  = C_{lm}^\text{M}(T) -\frac{1}{2(2l-1)}\frac{|\bm{X}|^2}{c^2}\frac{d^2}{dT^2}C_{lm}^\text{M}(T),
\end{equation}
\begin{equation}
\label{eq: mathcal Slm}
    \mathcal{S}_{lm}^\text{M}(T, |\bm{X}|)  = S_{lm}^\text{M}(T) -\frac{1}{2(2l-1)}\frac{|\bm{X}|^2}{c^2}\frac{d^2}{dT^2}S_{lm}^\text{M}(T).
\end{equation}
Here, the coefficients $C_{lm}^\text{M}(T)$ and $S_{lm}^\text{M}(T)$ are equivalent to the set of Mercury's mass multipole moments $\mathcal{M}^\text{M}_L$ and are related via time-dependent rotations to a set of approximately constant potential coefficients defined in a reference frame that co-rotates with Mercury. This transformation accounts for the planet’s rotation and allows the use of static gravitational field models in a rotating coordinate system, which represents a good approximation to the real time-dependent gravitational field of Mercury.

Following \cite{Soffel2003}, it can be shown that the second time derivative terms in Eqs.~(\ref{eq: mathcal Clm})--(\ref{eq: mathcal Slm}) can be safely neglected in the expansion of Mercury's local gravitational potential (Eq.~(\ref{eq: WM mathcal spher})).  

To give an explicit expression for the post-Newtonian part of $\Phi^a_\text{M}$ in terms of the multipole moments, we will make the following approximations: we will neglect all the higher spin moments $\mathcal{S}^\text{M}_L$ for $l>1$, along with all the mass multipole moments $\mathcal{M}^\text{M}_L$ for $l>2$; we will neglect all time derivatives of the multipole moments $\dot{\mathcal{M}}^\text{M}_L$, $\ddot{\mathcal{M}}^\text{M}_L$, $\dots$; we will discard terms which are quadratic in Mercury's quadrupole moments $\mathcal{M}^\text{M}_{ab}$ or bilinear in $\mathcal{M}^\text{M}_{ab}$ and $\mathcal{S}^\text{M}_c$. Moreover, following \cite{DSX1, DSX4}, we will take the origin of the local system to coincide with Mercury's center of mass. This implies that the dipole mass moment vanishes, i.e., $\mathcal{M}^\text{M}_a(T)=0$. 

With these assumptions, it is easy to show that the post-Newtonian part of $\Phi^a_\text{M}$ can be split as \cite{DSX4}
\begin{equation*}
    {[\Phi^a_\text{M}]_\text{PN}} = \Phi^a_\text{Schw} + \Phi^a_\text{LT} + \Phi^a_\text{RQ}+ \dots ,
\end{equation*}
where $ \Phi^a_\text{Schw}$ is the Schwarzschild acceleration, resulting from Mercury's mass monopole, $\Phi^a_\text{LT}$ is the Lense-Thirring acceleration, which arises from Mercury's spin vector $\bm{S}_\text{M}=\mathcal{S}^\text{M}_a$, and $\Phi^a_\text{RQ}$ is the relativistic acceleration which results from the quadrupole mass moments.

The Schwarzschild acceleration has the well-known expression
\begin{equation*}
   \! \Phi^a_\text{Schw}=\frac{GM_\text{M}}{c^2 |\bm{X}_\text{s}|^3}\! \left[ \! \left( 4\frac{GM_\text{M}}{|\bm{X}_\text{s}|}-|\bm{V}_\text{s}|^2\right)\! X^a_\text{s} + 4(\bm{X}_\text{s}\cdot \bm{V}_\text{s}) V^a_\text{s}\right]\! , 
\end{equation*}
where $\bm{V}_\text{s}=d\bm{X}_\text{s}(T)/dT$ is the velocity of the satellite.

The ``relativistic quadrupole" acceleration $\Phi^a_\text{RQ}$ reads
\begin{eqnarray*}
    \Phi^a_\text{RQ}&=& \frac{G\mathcal{M}^\text{M}_{bc}}{c^2}\left\{-2GM_\text{M} \left[\partial_{bc}\frac{1}{|\bm{X}_\text{s}|}\partial_a\frac{1}{|\bm{X}_\text{s}|} \right. \right. \\
    && \left. + \frac{1}{|\bm{X}_\text{s}|}\partial_{abc}\frac{1}{|\bm{X}_\text{s}|} \right] -2V^a_\text{s}V^d_\text{s} \partial_{bcd}\frac{1}{|\bm{X}_\text{s}|} \\
    && \left.  + \frac{1}{2} |\bm{V}_\text{s}|^2 \partial_{abc}\frac{1}{|\bm{X}_\text{s}|}\right\}. 
\end{eqnarray*}
The mass quadrupole moments can be expressed in terms of the spherical harmonics coefficients. Defining $\bm{M} = (\mathcal{M}^\text{M}_{ij})$, the matrix of the quadrupole moments expressed in the inertial frame, it holds that 
\begin{equation*}
    \bm{M} = \bm{P}^T \hat{\bm{M}} \bm{P},
\end{equation*}
where $\hat{\bm{M}} = (\hat{\mathcal{M}}^\text{M}_{ij})$ is the matrix of the quadrupole moments expressed in Mercury's body-fixed frame, and $\bm{P}(t)$ is the time-dependent matrix of transformation of coordinates from the inertial to the body fixed frame. 
The rotation matrix $\bm{P}(t)$ is given by
\begin{equation*}
    \bm{P}(t)=\mathcal{R}_3(\mathcal{W}^\text{M}(t))\mathcal{R}_1(\pi/2-\delta_0^\text{M})\mathcal{R}_3(\alpha_0^\text{M}+\pi/2),
\end{equation*}
where $\mathcal{R}_i(\psi)$ denotes the rotation matrix of angle $\psi$ around the $i$-th axis, i.e.: 
\begin{equation*}
   \!\! \mathcal{R}_1(\psi)\!=\!\! \begin{bmatrix}
        1 & 0 & 0 \\
        0 & \cos\psi & \sin\psi \\
        0 & - \sin\psi & \cos\psi\\
    \end{bmatrix}\!\!, \;
     \mathcal{R}_3(\psi)\!=\!\!\begin{bmatrix}
        \cos\psi & \sin\psi & 0\\
        - \sin\psi & \cos\psi & 0\\
        0 & 0 & 1
    \end{bmatrix}\!\!.
\end{equation*}
The angles $\alpha_0^\text{M}$ and $\delta_0^\text{M}$ are the right ascension and declination of Mercury's north pole, and the angle $\mathcal{W}^\text{M}$ specifies the ephemeris position of the prime meridian. The recommended values for such angles are reported in \cite{archinal}. 
Note that this formulation for Mercury's rotation is purely Newtonian, while a relativistic theory of Mercury's rotation should be employed for a consistent treatment (a relativistic modeling of rotational motion of extended bodies has been recently developed in \cite{relrotation}).

The matrix $\hat{\bm{M}}$ can be written in terms of the dimensionless spherical harmonics coefficients of order 2 as
\begin{equation*}
    \hat{\bm{M}} = M_\text{M}R_\text{M}^2 
        \begin{bmatrix}
            -\dfrac{1}{3}C_{20}^\text{M} +2 C_{22}^\text{M} & 2S_{22}^\text{M} & C_{21}^\text{M} \\[3pt]
             2S_{22}^\text{M} & -\dfrac{1}{3}C_{20}^\text{M} -2 C_{22}^\text{M} &  S_{21}^\text{M} \\[3pt]
             C_{21}^\text{M} & S_{21}^\text{M}  & \dfrac{2}{3}C_{20}^\text{M}\\
        \end{bmatrix}.
\end{equation*}
Such coefficients, expressed in the Mercury body-fixed frame, can be considered constants to the required level of accuracy.

Note that the trace of $\hat{\bm{M}}$ is zero; therefore, the zero trace property also holds for the matrix ${\bm{M}}$, since the two matrices are similar. 

Let $Y^a_\text{s}=\mathcal{M}^\text{M}_{ab}X^b_\text{s}$, then $\Phi^a_\text{RQ}$ can be written as sum of three terms: 
\begin{equation*}
\label{eq: phirq}
    \Phi^a_\text{RQ}=  \Phi^a_1 +\Phi^a_2 +\Phi^a_3,
\end{equation*}
where 
\begin{eqnarray*}
\label{eq: phi1}
    \Phi^a_1 &= &12\frac{GM_\text{M}}{|\bm{X}_\text{s}|^6}\frac{G}{c^2}\left( 3\frac{\left(\bm{Y}_\text{s}\cdot\bm{X}_\text{s} \right)}{|\bm{X}_\text{s}|^2} X^a_\text{s} - Y^a_\text{s}\right), \\
    \label{eq: phi2}
    \Phi^a_2 &=&\! \frac{6G}{c^2 |\bm{X}_\text{s}|^5} \! \left(\frac{5}{|\bm{X}_\text{s}|^2}(\bm{Y}_\text{s}\cdot \bm{X}_\text{s})(\bm{V}_\text{s}\cdot \bm{X}_\text{s}) - 2 (\bm{Y}_\text{s}\cdot \bm{V}_\text{s})\!\right) \!V^a_\text{s},\\
     \label{eq: phi3}
     \Phi^a_3 &=& \frac{3}{2} \frac{G}{c^2} \frac{|\bm{V}_\text{s}|^2}{|\bm{X}_\text{s}|^5}\left( 2Y^a_\text{s} - 5 \frac{X^a_\text{s}}{|\bm{X}_\text{s}|^2} (\bm{Y}_\text{s}\cdot \bm{X}_\text{s})\right).
\end{eqnarray*}

Finally, the Lense-Thirring acceleration has the expression 
\begin{equation*}
   \! \Phi_\text{LT}^a \!= \! \frac{2G}{c^2|\bm{X}_\text{s}|^3}\!\left[\left(\bm{V}_\text{s}\times\bm{S}_\text{M} \right)^a \! + \!\frac{3}{|\bm{X}_\text{s}|^2}\!\left(\bm{X}_\text{s}\cdot\bm{S}_\text{M}\right)\!\left( \bm{X}_\text{s}\!\times\!\bm{V}_\text{s}\right)^a \right]\!,
\end{equation*}
where the angular momentum vector $\bm{S}_\text{M}$ of Mercury can be computed as
\begin{equation*}
    \bm{S}_\text{M} = I_\text{M} \omega_\text{M} \bm{\hat{s}}_\text{M},
\end{equation*}
where $I_\text{M} \approx 0.353 M_\text{M} R_\text{M}^2$ is Mercury's polar moment of inertia,  $\omega_\text{M}$ is Mercury's spin angular velocity, and $\bm{\hat{s}}_\text{M}$ is the unit vector aligned with Mercury's spin axis and directed towards the north pole. Note that here we are neglecting precession, nutation, and polar motion, and we are assuming that the angular velocity vector is always directed towards the north pole. In the inertial frame, the vector $\hat{\bm{s}}_\text{M}$ can be written as 
\begin{equation*}
    \hat{\bm{s}}_\text{M}= \begin{pmatrix}
    \cos \alpha^\text{M}_0 \cos \delta^\text{M}_0 \\
    \sin \alpha^\text{M}_0 \cos \delta^\text{M}_0\\
    \sin \delta^\text{M}_0
\end{pmatrix}.
\end{equation*}
The constant 0.353 corresponds to the value of the moment of inertia divided by $M_\text{M} R_\text{M}^2$ obtained from the radio tracking data of the MESSENGER spacecraft \cite{smith_messenger}.

\subsection{Estimation of the norms of ${Q}_a$ and $\dot{{Q}}_a$}

\label{sec: Qa}

As stated above, the function $Q_a$ accounts for the deviation of the origin of the local frame from a geodesic of the external BCRS metric, due to the interaction of Mercury's multipole moments with the external bodies. As a comparison, in the case of the Earth, $Q_a$ is of the order of $3 \times 10^{-11}$~m/s$^2$, with the main contribution given by the gravitational field of the Moon \cite{brumbergkop1990}. 

In order to evaluate the order of magnitude of the function $Q_a$ for the case of Mercury, one can proceed as follows. First of all, it holds that (see, e.g., Eq.~(9.38) of \cite{relcelmechPPN}):
\begin{equation*}
    \ddot{{\mathcal{M}}}^\text{M}_a = \sum_{l=0}^{\infty} \frac{1}{l!} \mathcal{M}^\text{M}_L {Q}_{aL} + O(c^{-2}).
\end{equation*}
The definition of Mercury's center of mass implies that $\ddot{{\mathcal{M}}}^\text{M}_a=0$, from which it follows that
\begin{eqnarray}
\nonumber
   {Q}_a &=& - \frac{1}{M_\text{M}}  \sum_{l=2}^{\infty} \frac{1}{l!} \mathcal{M}^\text{M}_L {Q}_{aL} + O(c^{-2}) \\
   \label{eq: Qa}
   &&= -\frac{1}{2M_\text{M}} \mathcal{M}^\text{M}_{ij} {Q}_{aij} + \dots
\end{eqnarray}
where the dots indicate terms containing multipole moments for $l\geq 3$ and terms of order $O(c^{-2})$. Let $\bm{M} = (\mathcal{M}^\text{M}_{ij})$ be the matrix of the quadrupole moments expressed in the inertial frame.

Moreover, it holds that 
\begin{widetext}
\begin{equation*}
 {Q}_{ijk}=\overline{w}_{,ijk}(\bm{x}_\text{M}) = \sum_{\text{A}\neq \text{M}} 15 \frac{GM_\text{A} }{|\bm{r}_{\text{MA}}|^5}\left(\frac{1}{5}\delta_{ij}r_\text{MA}^k +\frac{1}{5}\delta_{ik}r_\text{MA}^j +\frac{1}{5}\delta_{jk}r_\text{MA}^i - \frac{r_\text{MA}^i r_\text{MA}^j r_\text{MA}^k}{|\bm{r}_\text{MA}|^2} \right), 
\end{equation*} 
\end{widetext}
and from the zero trace property of ${\bm{M}}$ it readily follows that 
\begin{equation*}
    \mathcal{M}^\text{M}_{ij}{Q}_{aij}= 15 \sum_{\text{A}\neq \text{M}} \frac{GM_\text{A}}{|\bm{r}_{\text{MA}}|^5} s_\text{A}^a,
\end{equation*}
with
\begin{equation*}
    s_\text{A}^a= \frac{2}{5}p_\text{A}^a - \frac{r_\text{MA}^a} {|\bm{r}_\text{MA}|^2} \left(r_\text{MA}^j p_\text{A}^j\right), 
\end{equation*}
and
\begin{equation*}
    p_\text{A}^i=\mathcal{M}^\text{M}_{ij} r_\text{MA}^j.
\end{equation*}
From a direct computation, it follows that 
\begin{equation*}
   |\bm{s}_\text{A}|^2 = \frac{4}{25} |\bm{p}_\text{A}|^2 + \frac{1}{5}\frac{\left(r_\text{MA}^i p_\text{A}^i\right)^2}{|\bm{r}_\text{MA}|^2}\leq \frac{9}{25} |\bm{p}_\text{A}|^2.
\end{equation*}
 Moreover, we have that 
\begin{equation*}
    |\bm{p}_\text{A}|^2 \leq ||\bm{M}||_2 \,|\bm{r}_\text{MA}|^2,
\end{equation*}
where $||\cdot||_2$ denotes the spectral norm of a matrix, equal to the maximum singular value of the matrix $\sigma_\text{max}$, i.e. $||\bm{A}||_2=\sigma_\text{max}(\bm{A})$ for any matrix $\bm{A}$.

From the definition of $\bm{M}$, one has $||\bm{M}||_2 = ||\hat{\bm{M}}||_2 = \sigma_\text{max}(\hat{\bm{M}})$. Therefore, 
\begin{equation}
    \label{eq: Qa stima}
    |\bm{Q}|\leq \frac{9}{2 M_\text{M}} \sigma_\text{max}(\hat{\bm{M}})\sum_{\text{A}\neq \text{M}} \frac{GM_\text{A}}{|\bm{r}_\text{MA}|^4}.
\end{equation}

Using the values given in the Mercury gravity solution HgM009 for the dimensionless spherical harmonic coefficients (see \cite{GENOVA2023115332}), one can compute $\sigma_\text{max}(\hat{\bm{M}})= M_\text{M}R_\text{M}^2\cdot 3.3528 \times 10^{-5}$. 
Inserting the corresponding values for the quantities on the right-hand side of Eq.~(\ref{eq: Qa stima}), we conclude that
\begin{equation}
\label{eq: Qa stima numeri}
    |\bm{Q}| \leq 2.66 \times 10^{-14} \; \text{m/s}^2,
\end{equation}
where, among all solar system bodies, the Sun gives the main contribution to the sum in Eq.~(\ref{eq: Qa stima}). We underline that Eq.~(\ref{eq: Qa stima numeri}) is obtained in the approximation where multipoles of order greater than 2 and terms of order $O(c^{-2})$ are discarded. Indeed, these terms are expected to be even smaller, so they can be safely neglected. 

Furthermore, in order to give an upper bound to the function $\dot{{Q}}_a= d{Q}_a/dT$, one can proceed by numerical differentiation. For instance, using Ridders' method \cite{numericalrecipes} over a one-year interval starting on the 1st of April 2027, we obtain 
\begin{equation}
\label{eq: Qadot stima}
    |\dot{\bm{Q}}| \leq 3.5 \times 10^{-20} \; \text{m/s}^3.
\end{equation}

\begin{figure}[h!]
    \centering
    \includegraphics[width=\linewidth]{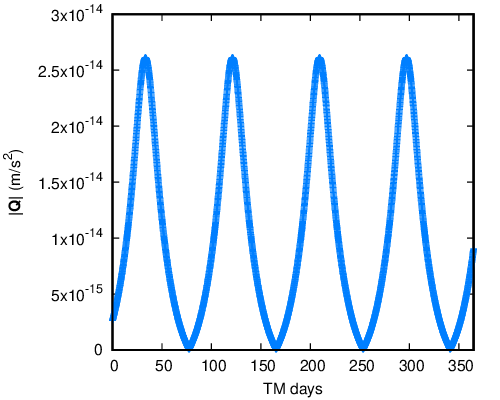}
\caption{\label{fig: Qa} Norm of ${Q}_a$ for Mercury during one year, starting on the 1st April 2027.}
\end{figure}

\begin{figure}[h!]
    \centering
    \includegraphics[width=\linewidth]{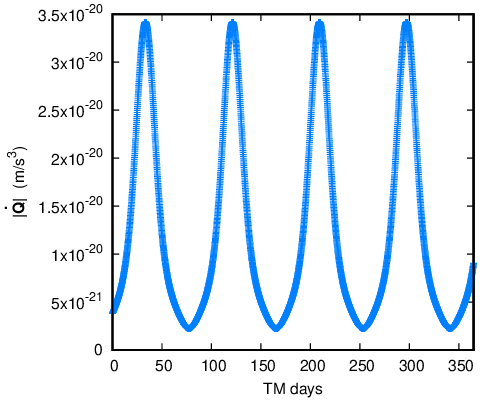}
\caption{\label{fig: Qadot} Norm of $\dot{Q}_a$ for Mercury during one year, starting on the 1st April 2027.}
\end{figure}

In Fig.~\ref{fig: Qa} and Fig.~\ref{fig: Qadot}, we show the norm of ${Q}_a$ for Mercury obtained from Eq.~(\ref{eq: Qa}) and the norm of $\dot{Q}_a$ obtained with numerical differentiation during one year, starting on the 1st of April 2027. We notice that the analytical estimate for the norm of ${Q}_a$ derived in Eq.~(\ref{eq: Qa stima numeri}) agrees with the numerical estimate. We remark that Fig.~\ref{fig: Qa} and Fig.~\ref{fig: Qadot} show the norms of vectors $Q_a$ and $\dot{Q}_a$, respectively. This explains why the curves in these Figures change in phase rather than out of phase, as it would be expected for the components of these vectors.

\bibliography{apssamp}

\end{document}